\documentclass[11pt]{article}

\usepackage[preprint]{acl}

\usepackage{times}
\usepackage{latexsym}

\usepackage[T1]{fontenc}

\usepackage[utf8]{inputenc}

\usepackage{microtype}

\usepackage{inconsolata}

\usepackage{graphicx}

\usepackage{booktabs,amssymb,algorithm,algorithmic, amsmath}

\title{Agentic Music Post-Production}

\author{
 \textbf{Jaywon Koo\textsuperscript{1}\thanks{Work done while intern at Sony}},
 \textbf{Junghyun Koo\textsuperscript{2}},
 \textbf{Marco A. Martínez-Ramírez\textsuperscript{2}},
 \textbf{Wei-Hsiang Liao\textsuperscript{3}},
\\
 \textbf{Seungheon Doh\textsuperscript{3}},
 \textbf{Qiyu Wu\textsuperscript{3}},
 \textbf{Vicente Ordonez\textsuperscript{1}},
 \textbf{Yuki Mitsufuji\textsuperscript{2,3}},
\\
\\
 \textsuperscript{1}Rice University,
 \textsuperscript{2}Sony Corporate Technology Center America, Inc.,
 \textsuperscript{3}Sony Group Corporation
\\
 \small{
   \textbf{Correspondence:} \href{mailto:jk125@rice.edu}{jk125@rice.edu}
 }
}

\newcommand{\method}{MixAgent}

\usepackage{placeins}
\usepackage{xcolor}
\usepackage{array}
\usepackage{xspace}
\usepackage{tikz}
\usepackage[most]{tcolorbox}
\usepackage{listings}
\usetikzlibrary{positioning,fit,backgrounds,arrows.meta}

\newcommand{\mixcritic}{MixCritic\xspace}

\newcommand{\gr}[1]{\textcolor{gray}{#1}}

\newcommand{\worse}[1]{\,{\tiny\textcolor{red!75!black}{#1}}}

\usepackage{booktabs}
\usepackage{graphicx}
\usepackage{xcolor}
\usepackage{colortbl}
\usepackage{tabularx}

\definecolor{mixmodelband}{RGB}{237,237,237}
\definecolor{mixoursband}{RGB}{246,243,255}
\definecolor{mixdeltabad}{RGB}{178,54,63}
\definecolor{mixdeltagood}{RGB}{28,118,75}
\definecolor{mixdeltazero}{RGB}{118,118,118}
\providecommand{\method}{MixAgent}

\newif\ifmixicons
\mixiconstrue
\DeclareRobustCommand{\mixice}{%
  \ifmixicons
    \raisebox{-0.15ex}{%
      \includegraphics[height=2.0ex]{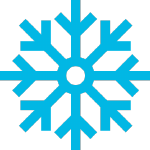}}%
  \fi}
\DeclareRobustCommand{\mixfire}{%
  \ifmixicons
    \raisebox{-0.15ex}{%
      \includegraphics[height=2.0ex]{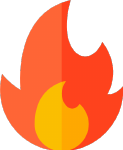}}%
  \fi}

\begin{document}

\maketitle
\begin{abstract}
We introduce \method{}, an agentic pipeline for music post-production that follows the workflow of audio engineers to build interpretable and editable mixing plans guided by a reference song. 
A grouping agent first routes related input tracks, such as drum recordings, into subgroups. A planning agent then uses retrieved examples to choose and order audio effects, such as volume control and reverb, for tracks, subgroups, and the full mix. To refine the mix, a detector identifies what needs correction, and a critic proposes numerical parameter adjustments from tracks to the full mix.
In our experiments, prompting recent multimodal models to listen and revise often worsens the mix, making reliable correction a central challenge.
We therefore build MixCritic, a dataset of 45,657 repair critiques derived from parameters fitted to professional mixes, and train the critic with supervised fine-tuning followed by preference optimization on rendered audio outcomes. On MedleyDB, learned refinement reduces error by 19\% relative to the initial mix, with preference optimization reducing it by 8.4\% over supervised training alone. Retrieved examples further improve effects selection, and a human listening test shows that \method{} gets higher quality ratings than mixes produced by Qwen3-Omni, Gemini 3.8 Flash and GPT-4o-audio.
\end{abstract}

\section{Introduction}
\label{intro}

Language-model agents use tools to write code~\citep{yang2024sweagent} and edit images~\citep{ye2026agentbanana}. In creative software, a key challenge is turning a high-level goal into precise settings. 
Noticing that ``the vocal is too quiet'' does not specify which control to move or by how much. \emph{Music mixing} combines separately recorded instruments and vocals (\emph{input tracks}) into a song by applying audio effects that control volume, frequency balance, and acoustic space.
Related tracks can also be processed together as \emph{subgroups}, such as all drum recordings, before processing the \emph{full mix}.
Settings interact, as louder drums can hide the vocal.
Because the desired sound can be difficult to describe in words, we study \emph{reference-guided mixing}, where a reference song provides an example.

\begin{figure}[t]
  \centering
  \includegraphics[width=\linewidth]{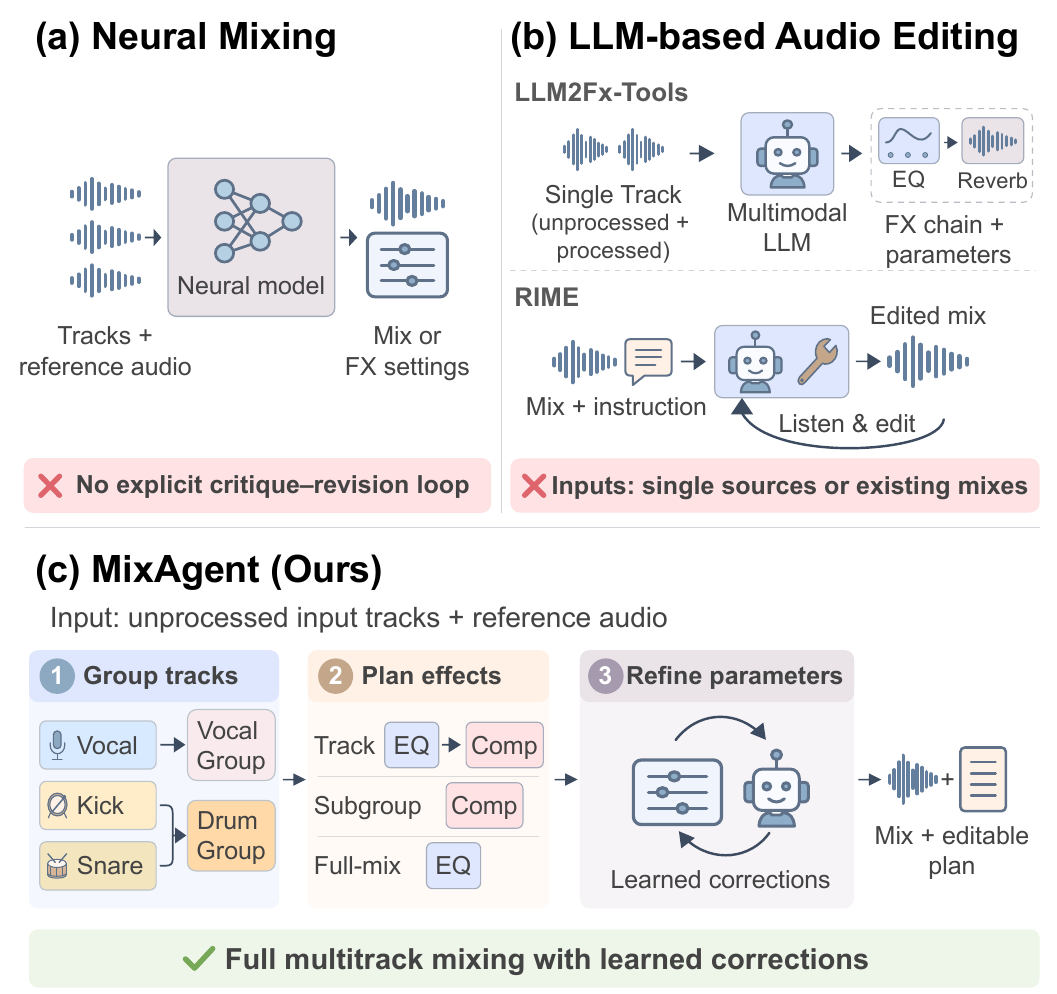}
  \caption{Comparison of music post-production workflows. 
  (a) Neural mixing systems predict final mixes or audio effects settings without explicit critique--revision loops. 
  (b) Prior LLM-based approaches focus on audio effects chain prediction and iterative mix refinement.
  (c) \method{} introduces an agentic framework that integrates hierarchical planning, execution, and iterative refinement into a unified music post-production pipeline.
  }
  \label{fig:teaser}
  \vspace{-0.15in}
\end{figure}

Existing systems address parts of this problem (Figure~\ref{fig:teaser}).
Neural and generative models produce competitive audio without exposing
audio effects settings~\citep{koo2023mixingstyletransfer,moliner2025megami},
while Diff-MST predicts editable parameters for fixed sequences of audio effects~\citep{vanka2024diffmst}. Prior LLM-based systems address narrower tasks.
LLM2Fx-Tools predicts audio effects chains for individual tracks~\citep{doh2026llm2fxtools}, and RIME edits an existing mix to follow instructions~\citep{schaffer2026rime}.
Building a mix from tracks requires choosing and ordering effects at the track, subgroup, and full-mix levels, setting their parameters, and identifying corrections without editing instructions.
Simply prompting an audio-language model to listen and revise does not consistently improve the mix.

\method{} builds an editable mixing plan and refines its parameters. A grouping agent organizes related tracks into subgroups. Guided by retrieved examples, a frozen planner chooses and orders audio effects at all three levels, then initializes their parameters. A learned detector selects \emph{what needs correction}, and a critic proposes \emph{how to fix it}. Refinement makes one pass at each level, from tracks to subgroups to the full mix, so later stages hear earlier corrections.
Because no known corpus pairs mixing critiques with parameter repairs, we build MixCritic, containing 45,657 critiques.
We train the critic with supervised fine-tuning (SFT), then direct preference optimization (DPO), favoring corrections toward the audio produced by the fitted parameters. Every effect, parameter, and correction remains explicit and editable.

The design also raises a broader question. \emph{What makes feedback useful for agents acting through numerical tools?} 
Our analyses show that retrieved examples improve audio effects selection, while learned refinement improves their parameters.
This refinement benefits from more specific feedback, with numerical corrections giving lower reconstruction error than perceptual or tool-specific advice.
With numerical feedback, DPO reduces this error by a further 8.4\% over SFT.
Beyond feedback, selecting correction targets and refining tracks before subgroups and the full mix further reduces error.

Training and objective evaluation use non-overlapping excerpts of the same song as inputs and references, allowing comparison against its professional mix. On held-out MedleyDB songs, \method{} achieves lowest reconstruction error and higher predicted quality than all tested multimodal LLM baselines. An 11-participant listening test with references from other songs within the same genre also gives \method{} higher quality ratings. Our contributions are:

\begin{itemize}
\setlength{\itemsep}{1pt}\setlength{\parskip}{0pt}
    \item To our knowledge, \method{} is the first LLM mixing agent to combine editable audio effects planning across tracks, subgroups, and the full mix with learned detection and correction of parameter errors.\footnote{Demo: \tiny\text{\url{https://github.com/SonyResearch/Agentic_Music_Post-Production}}} 
    \item We construct MixCritic, providing repair critiques in three feedback forms and preference pairs ranked by the resulting audio. 
    \item We show complementary reconstruction gains from retrieval and learned numerical refinement, with further improvements from selecting correction targets and refining tracks before subgroups and the full mix. 
\end{itemize}
\section{Preliminaries}
\label{sec:background}

\begin{figure*}[t]
  \centering
  \includegraphics[width=0.95\textwidth]{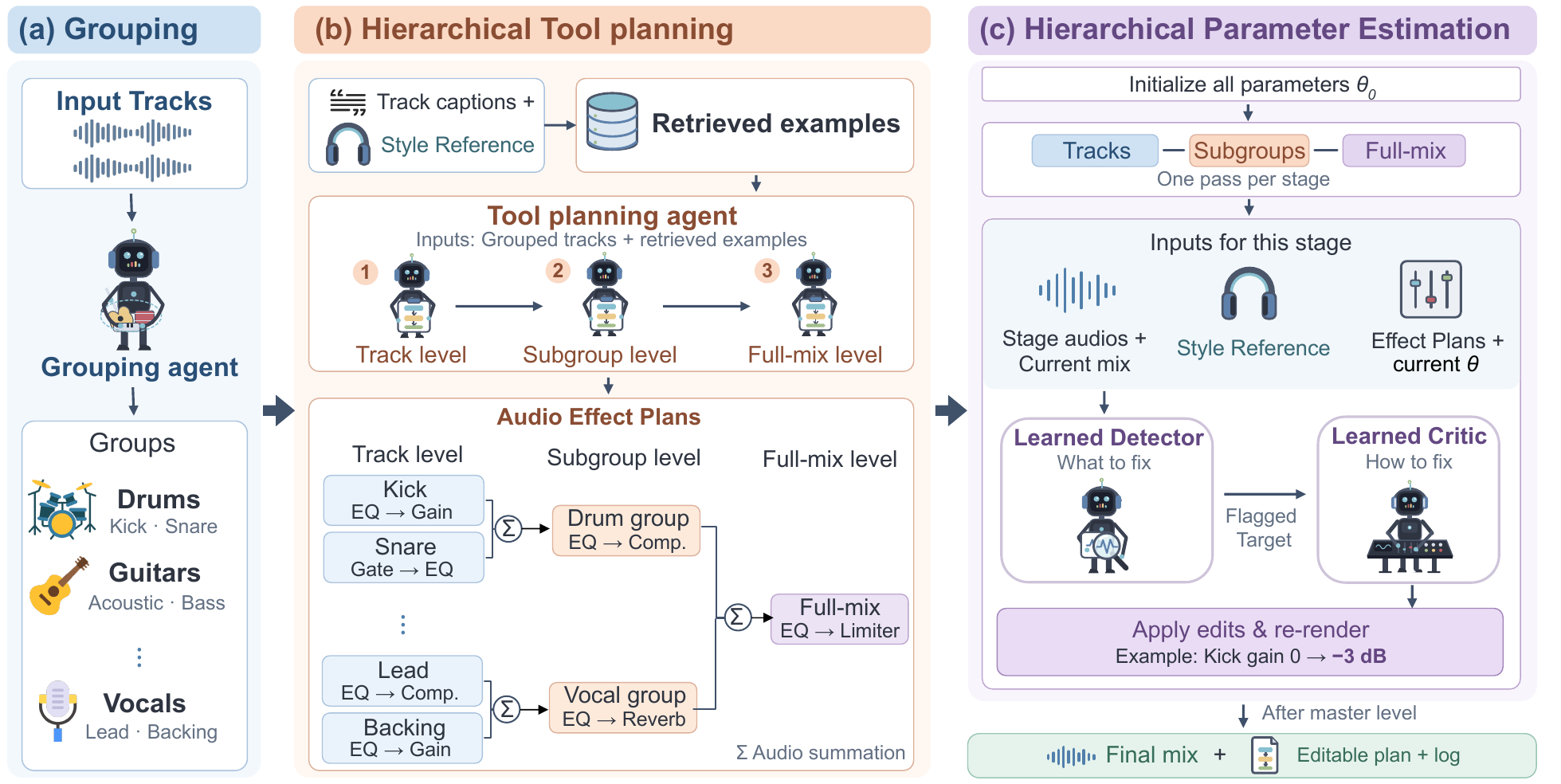}
  \caption{\textbf{Overview of MixAgent.}
  (a) The grouping agent assigns tracks to subgroups. (b) Using retrieved examples, the planner selects audio effects chains sequentially at the track, subgroup, and full-mix levels.
  (c) After parameter initialization, the detector flags elements for improvement, and the critic provides perceptual, tool-specific, or numerical feedback. Refinement proceeds from tracks to subgroups to the full mix.
}
  \label{fig:mixagent-overview}
  \vspace{-0.05in}
\end{figure*}

\paragraph{Music Post-production.}
The input consists of individual audio recordings, called \emph{input tracks}, such as a lead vocal, a bass guitar, or a single drum microphone. Mixing proceeds at three levels: individual tracks, \emph{subgroups} of related tracks summed and processed together (e.g., all drum tracks), and the \emph{full mix} formed by summing all subgroup outputs. 
Subgroup and full-mix processing correspond to subgroup buses and the master bus, respectively. The output after all processing is the \emph{final mix}. Mixing engineers shape the audio at each level with \emph{audio effects}, signal-processing tools controlled by numerical parameters. Typical audio effects include gain (volume), panning (left--right position), equalization (EQ, the balance of low and high frequencies), compression (reducing loudness fluctuations), and reverb (a sense of acoustic space). An ordered list of audio effects, such as EQ followed by compression, is an \emph{audio effects chain}.

\paragraph{Task.} We study reference-guided mixing. The input is a set of unprocessed input tracks $X$ with instrument labels, a library of available audio effects $\mathcal{T}$ (Appendix~\ref{app:effects}), and a \emph{style reference} $r$, which is a finished mix with the desired mixing style from a different song. 
LLM2Fx-Tools~\citep{doh2026llm2fxtools} predicts audio effects chains for individual tracks. We extend this task to multitrack mixing, producing a mixing plan $G$ and parameters $\theta$. The plan assigns the tracks $x_i\in X$ to subgroups, where $S_b$ is the set of tracks in subgroup $b$ (possibly a single track), and assigns an audio effect chain to each track, each subgroup, and the full mix, written $f_i$, $f_b$, and $f_m$ with parameters $\theta_i$, $\theta_b$, and $\theta_m$. Executing the plan, which we call \emph{rendering}, produces the output mix $\widehat y$ in three stages:
\begin{equation}
    \begin{aligned}
    y_i&=f_i(x_i;\theta_i) &&\text{(track)}\\
    y_b&=f_b\big(\textstyle\sum_{i\in S_b} y_i;\,\theta_b\big) &&\text{(subgroup)}\\
    \widehat y&=f_m\big(\textstyle\sum_{b} y_b;\,\theta_m\big) &&\text{(full mix)}
    \end{aligned}
    \label{eq:mixing-task}
\end{equation}
We denote the audio after each stage by $R_1(\theta)=\sum_{i\in S_b}y_i$, $R_2(\theta)=\sum_b y_b$, and $R_3(\theta)=\hat{y}$.

The agent must infer audio effects chains and settings to match the reference's mixing style rather than copy audio, since instrumentation, arrangement, or genre may differ. It must also account for interactions across tracks, subgroups, and the full mix.
Our datasets lack ground-truth mixes with references from other songs. We therefore use two non-overlapping excerpts of the same song for training and main evaluation (\S\ref{sec:exp-setup}).
The listening test uses references from other songs within the same genre (\S\ref{sec:exp-human}). \method{} keeps every audio effect and setting explicit, so users can inspect the plan, change values, and render the mix again. Audio is produced only by applying these audio effects to the input tracks, avoiding hallucinated musical content and artifacts from direct neural audio generation.

\section{\method{}}
\label{sec:method}

\method{} runs automatically and splits mixing into three decisions (Figure~\ref{fig:mixagent-overview}). It decides \emph{which audio effects} to use (\S\ref{sec:method-toolplan}), \emph{what to fix} in the current mix, and \emph{how to fix it} (\S\ref{sec:method-paramfill}). We describe how the components that make the last two decisions are trained (\S\ref{sec:method-training}).

\subsection{Planning Audio effects Chains with Retrieved Examples}
\label{sec:method-toolplan}

A \emph{grouping agent} first listens to the input tracks, reads their instrument labels, and assigns each track to a subgroup (e.g., kick, snare, and cymbals form a drum subgroup). A frozen audio-language \emph{tool-planning agent} then selects audio effects chains. 
Without examples, the planner tends to select short chains (\S~\ref{sec:exp-tool-planning}). To guide its choices, we use GRAFX~\citep{lee2024grafx}, a differentiable audio-processing library, to estimate example plans.
GRAFX fits audio effect parameters to approximate each database song's professional mix, then removes effects that barely change the rendered audio~\citep{lee2025mixinggraphs}.
Each database entry stores these \emph{pseudo-ground-truth} (pseudo-GT) plans and a caption of the rendered mix. For a new song, text retrieval matches automatically generated input-track captions to database captions to find similar instrumentation. Audio retrieval matches the reference to rendered database mixes for similar mixing styles. We provide three text-retrieved and one audio-retrieved plan, excluding all excerpts of the input song.
These examples show effect choices and order, but not parameter values. The planner selects track, subgroup, and full-mix chains in order, with each level guided by earlier choices.
The chains remain fixed.
Appendix~\ref{app:planning-details} and~\ref{app:prompts} provide details and prompts.

\begin{figure}[t]
  \centering
  \includegraphics[width=0.90\columnwidth]{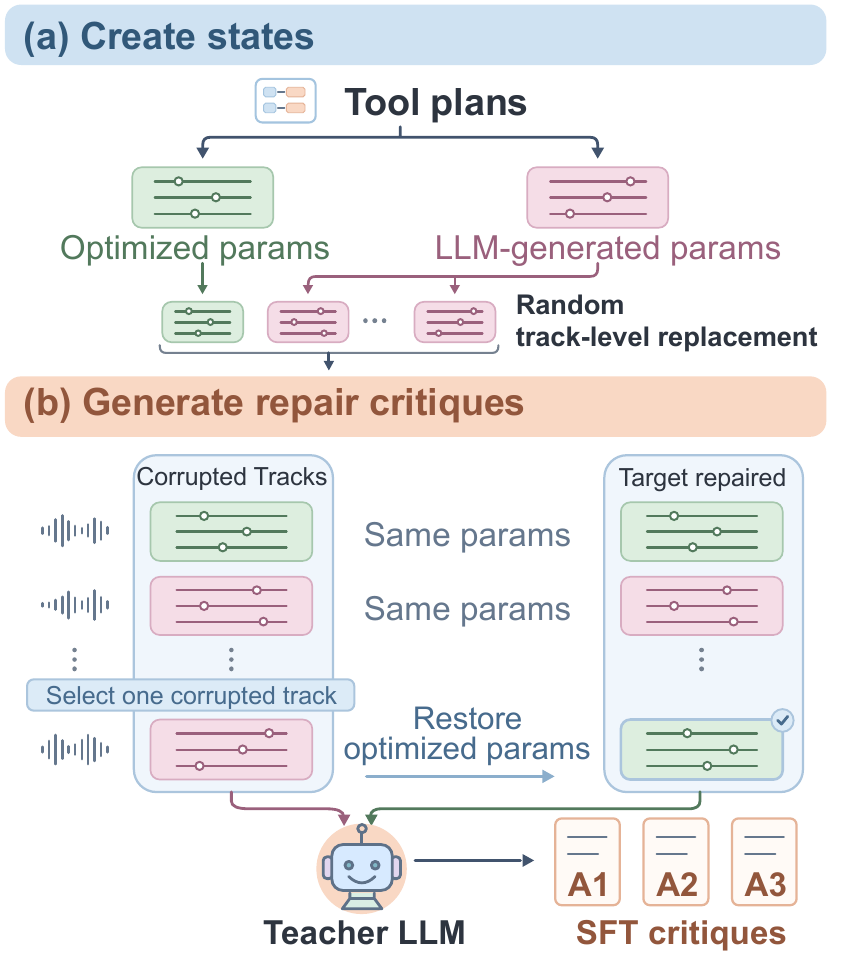}
  \vspace{-0.05in}
  \caption{
  Data generation for \mixcritic{}. (a) From a fixed tool plan, we obtain parameters through gradient-based optimization against the professional mix with GRAFX (green) and LLM generation (pink). Replacing the optimized parameters of randomly selected tracks with LLM outputs creates corrupted mixing states. (b) The teacher LLM compares the selected track’s corrupted and optimized parameters to generate A1–A3 critiques. The critic agent is trained to repair one corrupted track's parameters at a time.
  }
  \label{fig:mixcritic}
  \vspace{-0.15in}
\end{figure}

\subsection{Hierarchical Estimation}
\label{sec:method-paramfill}

The planner first sets the initial parameters of selected effect chains based on the reference audio.
\method{} then revises them \emph{bottom-up}, following common engineering practice~\citep{fox2024topdown}, first the tracks, then the subgroups, and finally the full mix (Figure~\ref{fig:mixagent-overview}). We call each track, subgroup, or the full mix an \emph{element} $e$, and two trained components carry out each revision.

\paragraph{Detector (what to fix).} At the track and subgroup stages, the trained \emph{detector} hears the processed tracks within a subgroup or the subgroup outputs, together with the reference $r$, and reads the current plan and parameters. For each $e$, it predicts the probability $p_e$ that an element needs correction and identifies audio effects that may need changes. The critic inspects elements with $p_e > \tau$, starting with the highest probability. The full mix is always inspected. Appendix~\ref{app:detector-arch} describes its architecture and training.

\paragraph{Critic (how to fix it).}
For each selected $e$, the audio-language \emph{critic} $\pi_\phi$ hears the audio of $e$ with its current audio effects, the current mix, and the reference. Its input $o_e$ also includes the detector's output and the audio effects chain and parameters of $e$. The critic proposes a correction $a_e$, and $\mathrm{Exec}$ applies it to the parameters of $e$:
\begin{equation}
    a_e\sim\pi_\phi(\cdot\mid o_e),\qquad
    \theta\leftarrow\mathrm{Exec}(a_e,\theta),
    \label{eq:critic}
\end{equation}
where $\mathrm{Exec}$ changes only the parameters of $e$. We compare three forms of feedback that describe the same repair (Table~\ref{tab:feedback-forms}). For perceptual advice (A1) and tool-specific advice (A2), it asks the planner to turn the advice into numbers. For numerical corrections (A3), it applies the proposed values directly. \method{} uses A3. 

Each \emph{refinement pass} calls the critic once per selected $e$, using audio from the pass's start. Then, applies corrections, re-renders the stage again, so the next stage hears the updated audio. We use one pass per stage, record every parameter change, and study additional passes (Appendix~\ref{app:refinement} and~\ref{app:iteration}).

\subsection{Learning from Repairs with \mixcritic{}}
\label{sec:method-training}
\label{sec:dataset-critic}

We build MixCritic, a dataset of corrupted mixing states, target repairs, and critiques (Figure~\ref{fig:mixcritic}).

\begin{table}[t]
\centering\small
\setlength{\tabcolsep}{3pt}
\begin{tabular}{@{}p{0.12\columnwidth}p{0.84\columnwidth}@{}}
\toprule
Form & Critic output (abridged) \\
\midrule
A1 & \textit{The vocal sounds heavily squashed and lifeless\ldots{} It needs to breathe naturally with smooth dynamics, a cleaner noise floor, and a touch of high-frequency air.} \\
A2 & {\footnotesize\raggedright
\{"tools": [\{"tool": "noisegate", "comment":
\textit{Raise the threshold significantly so it only engages on actual noise, and reduce the wet mix to make the processing transparent and eliminate pumping artifacts.}
\}
]\}
\par} \\
A3 & {\footnotesize\raggedright\ttfamily
\{"tool": "noisegate", "parameters":
\{"log\_threshold": 1.28, "log\_ratio": 1.04,
"log\_knee": 2.45, "z\_alpha\_pre": 4.35\},
"drywet": 0.599\} \textellipsis
\par} \\
\bottomrule
\end{tabular}
\caption{Critic output examples. A1 provides high-level semantic feedback, A2 names tools and revision directions, and A3 provides explicit numerical parameter values.} 
\label{tab:feedback-forms}
\vspace{-0.1in}
\end{table}

\paragraph{Corrupted states.}
For each of the 165 MedleyDB training examples, we fit the parameters of the agent's selected audio effects chains to the professional mix using GRAFX. This gives pseudo-GT parameters $\theta^\ast$ and rendered audio $y_k^\ast$ at each stage $k$. Qwen3-30B Thinking also predicts parameters for the same chains. We create a corrupted state $\tilde{\theta}$ by replacing the fitted parameters of randomly selected tracks with these predictions, then render the audio. The input tracks, audio effects chains, and pseudo-GT targets remain fixed.

\begin{table*}[!t]
\centering
\begingroup
\footnotesize
\setlength{\tabcolsep}{3pt}
\renewcommand{\arraystretch}{1.13}
\resizebox{1.0\linewidth}{!}{
\begin{tabularx}{\textwidth}{@{}>{\raggedright\arraybackslash}X*{10}{c}@{}}
\toprule
& \multicolumn{5}{c}{\textbf{MedleyDB}}
& \multicolumn{5}{c}{\textbf{AlbumDB}} \\
\cmidrule(lr){2-6}\cmidrule(l){7-11}
& Reconst. & \multicolumn{2}{c}{Quality} & \multicolumn{2}{c}{Similarity}
& Reconst. & \multicolumn{2}{c}{Quality} & \multicolumn{2}{c}{Similarity} \\
\cmidrule(lr){2-2}\cmidrule(lr){3-4}\cmidrule(lr){5-6}
\cmidrule(lr){7-7}\cmidrule(lr){8-9}\cmidrule(l){10-11}
Method & MSS $\downarrow$ & TuneJury $\uparrow$ & AAPQ $\uparrow$
& FxEnc++ $\uparrow$ & AF $\downarrow$
& MSS $\downarrow$ & TuneJury $\uparrow$ & AAPQ $\uparrow$
& FxEnc++ $\uparrow$ & AF $\downarrow$ \\
\midrule
Random & 7.04 & -2.18 & 5.96 & 0.51 & 8.7e4
 & 7.90 & -2.26 & 5.84 & 0.46 & 401.6 \\
\midrule
\rowcolor{mixmodelband}
\multicolumn{11}{c}{\strut\mixice\enspace\textit{Qwen3-Omni}} \\
Single-pass & 3.73 & -0.55 & 6.38 & 0.65 & 0.36
 & 4.34 & 0.44 & 6.44 & 0.78 & 0.14 \\
Self-Refine & 4.57 & -1.22 & 5.78 & 0.63 & 2.86
 & 4.50 & 0.17 & 6.09 & 0.76 & 0.21 \\
Self-Critic & 4.79 & -0.77 & 5.85 & 0.63 & 8.97
 & 4.55 & 0.23 & 6.06 & 0.77 & 0.10 \\
\midrule
\rowcolor{mixmodelband}
\multicolumn{11}{c}{\strut\mixice\enspace\textit{Gemini 3.8 Flash}} \\
Single-pass & 3.77 & \underline{-0.43} & \underline{6.52} & 0.68 & 0.20
 & 4.62 & \textbf{0.70} & \underline{6.70} & 0.79 & 0.08 \\
Self-Refine & 3.69 & \underline{-0.43} & 6.30 & \textbf{0.75} & \textbf{0.16}
 & 4.89 & \underline{0.54} & 6.21 & 0.78 & \underline{0.07} \\
Self-Critic & 3.73 & -0.50 & 5.67 & 0.71 & \underline{0.18}
 & 4.90 & 0.41 & 5.62 & 0.77 & 0.09 \\
\midrule
\rowcolor{mixmodelband}
\multicolumn{11}{c}{\strut\mixice\enspace\textit{GPT-4o-audio}} \\
Single-pass & \underline{3.67} & -0.47 & 6.47 & 0.66 & 0.29
 & 3.84 & 0.26 & 6.31 & \textbf{0.81} & \textbf{0.06} \\
Self-Refine & 3.95 & -0.79 & 6.20 & 0.64 & 0.30
 & \textbf{3.77} & 0.36 & 6.04 & \underline{0.80} & 0.08 \\
Self-Critic & 4.53 & -1.55 & 5.55 & 0.63 & 0.28
 & 4.91 & -0.40 & 6.25 & 0.77 & 0.10 \\
\midrule
\rowcolor{mixoursband}
\mixfire\enspace\textbf{\method{}} & \textbf{2.83} & \textbf{0.21} & \textbf{7.03} & \underline{0.73} & 0.20
 & \underline{3.81} & \textbf{0.70} & \textbf{7.41} & 0.79 & 0.11 \\
\bottomrule
\end{tabularx}
}
\endgroup
\caption{Results on MedleyDB and AlbumDB. Bold and underline mark the best
and second-best reported scores across all methods for each metric and
dataset.
\ifmixicons
\mixice\ denotes off-the-shelf LLM baselines without task-specific training; \mixfire\ marks \method{} with a trained detector and critic.
\fi
Single-pass uses no audio-feedback revision.}
\label{tab:main}
\vspace{-0.1in}
\end{table*}

\paragraph{Repair score.}
We score a correction $a$ by how much it reduces the difference between the rendered audio and the pseudo-GT audio.
\begin{equation}
    \begin{aligned}
    s(a)={}&\mathrm{MRS}\big(\mathcal{R}_k(\tilde\theta),y^\ast_k\big)\\
    &-\mathrm{MRS}\big(\mathcal{R}_k(\mathrm{Exec}(a,\tilde\theta)),y^\ast_k\big),
    \end{aligned}
    \label{eq:repair-score}
\end{equation}
where MRS measures audio distance~\citep{steinmetz2020auraloss}. Positive scores mean improvement toward the stage-$k$ pseudo-GT audio $y^\ast_k$.

\paragraph{Supervised critiques.}
For each corrupted state, we select one corrupted track and define a target correction $a^\ast$ that restores its pseudo-GT parameters. The other tracks keep their current settings.
A larger teacher LLM (Qwen3.6-35B) receives the corrupted state and the target track's LLM-generated and fitted parameters, and writes critiques in forms A1--A3, each with a step-by-step rationale.
We keep a critique if its target repair satisfies $s(a^\ast)>\delta_0$, leaving 45{,}657 examples. The detector is trained on the same states, plus states with random parameters, to predict which tracks, subgroups, and audio effects were corrupted (class-weighted binary cross-entropy). Pseudo-GT parameters are used only to build \mixcritic{} and never at inference.

\paragraph{Supervised fine-tuning (SFT).}
The critic uses a Qwen3.5-4B language model~\citep{qwen2026qwen35} and a frozen RLAT encoder trained to represent audio effects~\citep{lee2026audiotransformations}. A trainable audio adapter connects them. 
Following LLM2Fx-Tools~\citep{doh2026llm2fxtools}, we first train the adapter alone and then train it jointly with low-rank adaptation (LoRA) weights. The loss is $\mathcal{L}_{\mathrm{SFT}}=\mathcal{L}_{\mathrm{CE}}+0.3\,\mathcal{L}_{\mathrm{NTL}}$, which adds to token cross-entropy a number-token loss that penalizes numerically distant predictions~\citep{zausinger2025ntl}.

\paragraph{Preference training on audio outcomes (DPO).} 
SFT learns one target correction, but other corrections may also improve the mix. For each $e$ selected by the detector in a training state, the SFT critic samples eight candidate critiques. We apply each candidate to the same starting parameters, render the result, and normalize its repair score as $\hat{s}(a)=s(a)/s(a^\ast)$. The highest-scoring candidate is paired with each of the two lowest-scoring candidates when their normalized score gap is at least $\delta$. About 30\% of pairs come from detector false alarms, where $e$ was not corrupted and no target repair exists. For these, we normalize by the initial audio error $\mathrm{MRS}(\mathcal{R}_k(\tilde{\theta}),y_k^\ast)$ instead. Starting from the SFT critic, direct preference optimization (DPO)~\citep{rafailov2023dpo} favors higher-scoring corrections using the frozen SFT critic as reference. Only the audio adapter and LoRA weights are updated (learning rate of $5\times10^{-6}$, $\beta=0.1$, and 24 pairs per batch). 
Appendix~\ref{app:critic-training} gives the critic architecture and training settings, and Appendix~\ref{app:mixcritic} gives \mixcritic{} statistics.

\section{Experiments}
\label{sec:experiments}

\subsection{Setup}
\label{sec:exp-setup}
\label{sec:dataset}

\paragraph{Data.}
We build the task from MedleyDB~\citep{bittner2014medleydb}, using its unprocessed recordings as unprocessed input tracks and its professional mixes as targets. 
Our datasets lack ground-truth mixes made to match references from other songs. 
We therefore sample two non-overlapping 15-second windows from each song section: one provides the unprocessed input tracks and the \emph{ground-truth} mix (GT), for supervision and evaluation, and the other provides the style reference. 
This preserves a similar mixing style without providing the target audio as the reference. At evaluation, GT is used only for scoring. 
This gives 185 clips from 123 songs (Appendix~\ref{app:data}). We hold out 20 clips for evaluation, with artists disjoint from training. For cross-corpus transfer, we build AlbumDB clips~\citep{mckenzie2026albumdb} identically.

\paragraph{Baselines.}
We evaluate Qwen3-Omni~\citep{xu2025qwen3omni}, Gemini 3.8 Flash~\citep{google2026gemini38flash}, and GPT-4o-audio~\mbox{\citep{hurst2024gpt4o}} under three settings (Table~\ref{tab:main}). \textit{Single-pass} generates a mixing plan and its parameters without revision. \textit{Self-Refine} listens to its current mix and revises it directly.
\textit{Self-Critic} first writes a critique of the mix, then revises it using that critique. Both adapt \citet{madaan2023selfrefine} and iterate three rounds.
All settings use the same unprocessed input tracks, style reference, and list of available audio effects as \method{}.
A random baseline samples audio effects and parameters.

\paragraph{Metrics.}
We evaluate reconstruction, audio quality, and mixing-style similarity. Multi-Scale Spectral (MSS)~\citep{engel2020ddsp} compares the frequency content of the output and GT over time at several resolutions, using the sum and difference of the left and right channels. Multi-Resolution STFT (MRS)~\citep{yamamoto2020parallel}, used in training and analyses, makes this comparison for each channel. TuneJury~\citep{kim2026tunejury} predicts listener preference, and Audiobox~\citep{tjandra2025audiobox} predicts production quality (AAPQ). Neither uses GT or the style reference.
For similarity to GT, FxEnc++~\citep{yeh2025fxencoder} compares learned audio effects representations, while Audio Feature loss (AF)~\citep{vanka2024diffmst} compares hand-crafted features of volume, loudness fluctuations, frequency balance, and left--right spread. We omit RLAT similarity because we train on its representations.

\subsection{Main Results}
\label{sec:exp-main}

On MedleyDB, \method{} achieves the lowest MSS (2.83 vs.\ 3.67 for single-pass GPT-4o-audio) and the highest TuneJury and Audiobox PQ scores (Table~\ref{tab:main}). Gemini 3.8 Flash with Self-Refine scores best on FxEnc++ and AF. Results on AlbumDB are mixed. \method{} has the highest Audiobox PQ and ties single-pass Gemini on TuneJury, while its MSS is second best (3.81 vs.\ 3.77 for GPT-4o-audio with Self-Refine). Single-pass GPT-4o-audio scores best on FxEnc++ and AF.

Prompted revision does not consistently improve on single-pass mixing. Self-Critic yields higher MSS than Self-Refine for every model on both datasets. In \method{}, learned refinement lowers MSS from 3.70 to 2.83 with fixed audio effects chains (\S5). Appendix~\ref{app:dedicated} compares \method{} with dedicated mixing models (Diff-MST, MST, and StemFX). \method{} has the lowest MedleyDB MSS, although MST and StemFX use tracks extracted from the professional mix.

\begin{table*}[!t]
\centering
\begingroup
\footnotesize
\setlength{\tabcolsep}{6pt}
\renewcommand{\arraystretch}{1.13}
\newcommand{\mixoracle}[1]{\textcolor{black!60}{#1}}
\begin{tabularx}{\textwidth}{>{\raggedright\arraybackslash}X*{6}{c}}
\toprule
& \multicolumn{3}{c}{\textbf{Final output}}
& \multicolumn{3}{c}{\textbf{Applied revisions}} \\
\cmidrule(lr){2-4}\cmidrule(l){5-7}
Configuration & MRS $\downarrow$ & TuneJury $\uparrow$ & FxEnc++ $\uparrow$
& Improved (\%) $\uparrow$ & Mean $\Delta$MRS $\uparrow$ & Count \\
\midrule
No refinement (planner only) & 2.17 & -0.72 & 0.71
& \textemdash & \textemdash & \textemdash \\
\midrule
\rowcolor{mixmodelband}
\multicolumn{7}{c}{\strut\textit{Oracle feedback and parameter fitting (GT access)}} \\
\mixoracle{Oracle A1} & \mixoracle{1.87} & \mixoracle{-0.41} & \mixoracle{0.78}
& \mixoracle{53.6} & \mixoracle{+0.0331} & \mixoracle{250} \\
\mixoracle{Oracle A2} & \mixoracle{1.78} & \mixoracle{-0.44} & \mixoracle{0.80}
& \mixoracle{60.9} & \mixoracle{+0.0374} & \mixoracle{243} \\
\mixoracle{Oracle A3} & \mixoracle{0.90} & \mixoracle{0.51} & \mixoracle{0.93}
& \mixoracle{\textemdash} & \mixoracle{\textemdash} & \mixoracle{\textemdash} \\
\midrule
\rowcolor{mixmodelband}
\multicolumn{7}{c}{\strut\textit{A1: Perceptual advice (converted to settings by the planner)}} \\
\mixice\enspace Qwen3-Omni & 2.26 & -0.83 & 0.68 & 41.9 & -0.0212 & 117 \\
\mixfire\enspace SFT & 2.08 & -0.47 & 0.71 & 50.6 & +0.0050 & 245 \\
\mixfire\enspace SFT+DPO & 2.09 & -0.56 & 0.69 & 56.3 & +0.0026 & 245 \\
\midrule
\rowcolor{mixmodelband}
\multicolumn{7}{c}{\strut\textit{A2: Tool-specific advice (converted to settings by the planner)}} \\
\mixice\enspace Qwen3-Omni & 2.44 & -0.77 & 0.66 & 48.7 & -0.0079 & 113 \\
\mixfire\enspace SFT & 2.24 & -0.37 & \textbf{0.74} & 54.0 & +0.0045 & 252 \\
\mixfire\enspace SFT+DPO & 1.98 & -0.57 & 0.71 & 60.0 & +0.0220 & 215 \\
\midrule
\rowcolor{mixmodelband}
\multicolumn{7}{c}{\strut\textit{A3: Numerical corrections (applied directly)}} \\
\mixice\enspace Qwen3-Omni & 2.18 & -0.75 & 0.69 & 52.2 & -0.0076 & 113 \\
\mixfire\enspace SFT & \underline{1.91} & \underline{-0.30} & 0.70
& \underline{64.7} & \underline{+0.0221} & 238 \\
\rowcolor{mixoursband}
\mixfire\enspace\textbf{SFT+DPO (\method{})} & \textbf{1.75} & \textbf{0.21} & \underline{0.73}
& \textbf{66.8} & \textbf{+0.0398} & 241 \\
\bottomrule
\end{tabularx}
\endgroup
\caption{Critic training and feedback form on MedleyDB. Improved is the percentage of revisions that lower full-mix MRS, Mean $\Delta$MRS is the average reduction per revision, and  Count is the number of corrections
applied across all clips.
\ifmixicons
\mixice\ denotes prompted Qwen3-Omni critic without fine-tuning and
\mixfire\ marks trained detector and critic (trained critics use Qwen3.5-4B).
\fi
Gray oracle entries use GT.}
\label{tab:parameter_filling}
\vspace{-0.1in}
\end{table*}

\begin{figure}[!t]
  \centering
  \includegraphics[width=\linewidth]{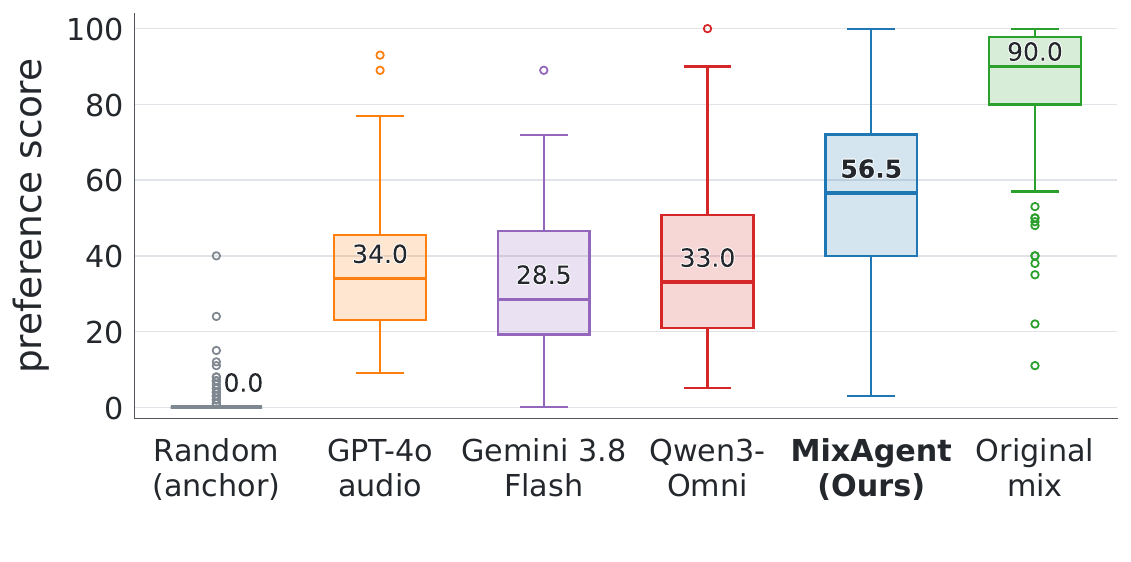}
  \vspace{-0.4in}
  \caption{Human evaluation of mixing quality.
  }
  \vspace{-0.15in}
  \label{fig:mixagent-mushra}
\end{figure}

\subsection{Human Listening Test}
\label{sec:exp-human}

\paragraph{Human Listening Test.}
We conducted a multi-stimulus listening test to evaluate perceptual music production quality in a practical reference-based mixing scenario. Input tracks were paired with a reference mix from a different song within the same genre.
Eleven participants completed 10 trials each, evaluating the production quality of mixes by \method, three LLM baselines, a randomly processed anchor, and the original human mix on a 0--100 scale.
Figure~\ref{fig:mixagent-mushra} shows that three LLM baselines exhibit largely overlapping score distributions, with no statistically significant differences among them. In contrast, MixAgent achieves significantly higher scores than all three baselines ($p < 0.001$, Holm-corrected paired t-tests), while a substantial perceptual gap remains relative to the original human mixes.

\section{What Makes Feedback Useful?}
\label{sec:analysis}
\label{sec:exp-ablations}

We examine what makes learned refinement effective.
Protocols are in Appendix~\ref{app:ablation-protocol}.

\subsection{Critic Training and Feedback Form}
\label{sec:exp-critic}

Table~\ref{tab:parameter_filling} compares critic training and feedback forms with fixed audio effect chains and detector.
Untrained critics make harmful corrections. Because our critic has a trained audio adapter, we use prompted Qwen3-Omni as the untrained critic. Its corrections increase error on average across all three forms, with only 42--52\% improving the mix. Final MRS is 2.18 to 2.44, versus 2.17 for the planner alone and 1.75 for trained A3 refinement. The base model's judgments are unreliable for correction.
Appendix~\ref{app:stage-error} provides stage-wise errors.

Preference training improves numerical feedback (A3) under the same detector, planner, and execution. MRS falls from 1.91 to 1.75 (8.4\%), mean improvement per correction nearly doubles (+0.022 to +0.040), and success rises from 64.7\% to 66.8\%. It also improves A2 (2.24 to 1.98), but not A1 (2.08 to 2.09), suggesting preference training benefits most from specific feedback.

After SFT+DPO, error falls with more specific feedback, from 2.09 MRS for perceptual advice to 1.98 for tool-specific advice and 1.75 for numerical corrections.
Without GT access, trained A3 also outperforms oracle A1/A2 advice (1.87/1.78) on MRS and TuneJury, suggesting that converting verbal advice into settings remains difficult. GT-fitting the same plans reaches 0.90 MRS, leaving substantial room for better parameter estimation.

\subsection{Choosing Audio effects with Retrieved Examples}
\label{sec:exp-tool-planning}

To isolate audio effects selection, Table~\ref{tab:tool-planning-main} fits each plan's parameters to GT on 100 MedleyDB clips and measure what the chosen audio effects can achieve. Retrieved examples improve the plans. The fitted MRS drops from 2.13 without examples to 1.08 with three text-retrieved and one audio-retrieved example, while random examples reach only 1.71. Text retrieval provides most of the gain. Retrieval also leads the planner to select more tools per track (3.8 vs. 1.6), which accounts for part of this gain.
Appendix~\ref{app:planning} reports audio effects chain F1, while Appendix~\ref{app:console-examples} shows \method{} plans with more engineer-like track grouping and effect ordering than GRAFX.

\subsection{Component Ablations}
\label{sec:exp-schedule}

Retrieval and learned refinement are complementary. We run the hierarchical planner with and without retrieved examples, each with and without learned refinement (Table~\ref{tab:components}). Retrieval lowers MRS by 0.64, both without refinement (2.81 to 2.17) and with it (2.39 to 1.75), and refinement lowers MRS by 0.42 in either case. MSS likewise improves by 0.82 from retrieval and 0.87 from refinement. Retrieval improves the choice of audio effects, and refinement improves their parameters, with additive gains.
We compare \method{}'s bottom-up order with a top-down order (full mix, then subgroups, then tracks) and with a variant where corrections are computed from the initial audio and applied at once. Bottom-up refinement gives the lowest MRS (1.75, vs.\ 1.82 and 1.88) and MSS (Table~\ref{tab:components}), suggesting that later corrections benefit from hearing earlier ones. 
With the critic fixed, detector selection yields lower MRS (1.75) than inspecting all tracks and subgroups (1.87) or an equally sized random subset (1.81), and the lowest MSS. Selecting appropriate targets matters more than inspecting additional elements.  Appendix~\ref{app:detector} reports detector accuracy.

\begin{table}[t]
\centering\small
\setlength{\tabcolsep}{3pt}
\begin{tabular*}{\columnwidth}{@{\extracolsep{\fill}}lccc@{}}
\toprule
Examples & MRS $\downarrow$ & FxEnc++ $\uparrow$ & Tools/track \\
\midrule
No examples & 2.13 & 0.70 & 1.55 \\
Random (3) & 1.71 & 0.75 & 1.74 \\
Text (3) & 1.10 & 0.94 & 3.75 \\
Text (3) + audio (1) & 1.08 & 0.94 & 3.77 \\
\midrule
GRAFX & 0.89 & 0.98 & 4.38 \\
\bottomrule
\end{tabular*}
\caption{Audio effects chain planning on 100 randomly selected MedleyDB clips.
Parentheses denote number of retrieved in-context examples. }
\label{tab:tool-planning-main}
\vspace{-0.1in}
\end{table}

\subsection{An Inspectable Output}
\label{sec:exp-interpretability}

Each step of \method{} acts on named audio effects and parameters, so its output records how the mix was made. On the evaluation set, the critic applies 241 corrections (about 12 per clip), each a change to a named parameter, and two thirds of them reduce the error of the full mix. Users can trace, revert, or adjust edits unavailable in audio-only outputs.
\section{Related Work}
\label{sec:related}

\paragraph{Agents with tools and learned feedback.} LLM agents call tools~\citep{yao2023react,schick2023toolformer} and refine their outputs with self-generated feedback~\citep{madaan2023selfrefine, shinn2023reflexion}, but self-correction without an external signal often fails~\citep{huang2024selfcorrect}, which motivates trained correctors~\citep{welleck2023selfcorrect}. In music production, LLM2Fx predicts audio effects parameters from text~\citep{doh2025llm2fx}, LLM2Fx-Tools recovers the audio effect chain of a single track through tool calls~\citep{doh2026llm2fxtools}, and MixAssist gives mixing advice in conversation~\citep{clemens2025mixassist}. RIME listens between tool calls and edit finished mixes from synthetic instruction--edit pairs~\citep{schaffer2026rime}. \method{} instead builds mixes from unprocessed input tracks, and its detector and critic are trained on whether each correction improves the rendered audio.

\paragraph{Automatic mixing.}
Explicit-processing methods estimate differentiable audio-effect parameters~\citep{steinmetz2021dmc,steinmetz2022deepafxst,vanka2024diffmst,steinmetz2024stito,koo2025itomaster} or variable-length chains~\citep{cheng2026stemfx}, but most use fixed consoles without selecting effects or grouping tracks. GRAFX fits graphs to known targets~\citep{lee2024grafx,lee2025mixinggraphs}, supplying our retrieval plans, training data, and oracles. Neural approaches include end-to-end mixing~\citep{martinezramirez2022fxnorm}, style transfer~\citep{koo2023mixingstyletransfer}, and diffusion or flow-based generation~\citep{levy2023controllable,melechovsky2025sonicmaster,moliner2025megami,yeh2026sequentialmixing}. Diff2Mix instead decodes generated embeddings into fixed-console parameters~\citep{zong2026diff2mix}. Direct audio generation can introduce artifacts and obscure audio effect settings~\citep{melechovsky2025sonicmaster,moliner2025megami}. \method{} combines open-ended chains, subgroups, and editable parameters with detection and correction learned from rendered audio.

\begin{table}[t]
\centering\footnotesize
\setlength{\tabcolsep}{4pt}
\resizebox{1.0\linewidth}{!}{
\begin{tabular}{@{}lcccc@{}}
\toprule
Configuration & MRS$\downarrow$ & MSS$\downarrow$
& TuneJury$\uparrow$ & FxEnc++$\uparrow$ \\
\midrule
\method{} (full)
& \textbf{1.75}
& \textbf{2.83}
& 0.21
& \textbf{0.73} \\
\midrule
\multicolumn{5}{@{}l}{\textit{Retrieval and learned refinement}} \\
\quad w/o learned refinement
& 2.17\worse{$+$0.42}
& 3.70\worse{$+$0.87}
& $-0.72$\worse{$-$0.93}
& 0.71\worse{$-$0.02} \\
\quad w/o retrieval
& 2.39\worse{$+$0.64}
& 3.65\worse{$+$0.82}
& \textbf{0.28}{\scriptsize\,\textcolor{green!50!black}{$+$0.07}}
& 0.70\worse{$-$0.03} \\
\quad w/o both
& 2.81\worse{$+$1.06}
& 4.52\worse{$+$1.69}
& $-0.82$\worse{$-$1.03}
& 0.65\worse{$-$0.08} \\
\midrule
\multicolumn{5}{@{}l}{\textit{Refinement order}} \\
\quad top-down (full mix first)
& 1.82\worse{$+$0.07}
& 3.07\worse{$+$0.24}
& 0.07\worse{$-$0.14}
& 0.69\worse{$-$0.04} \\
\quad all corrections at once
& 1.88\worse{$+$0.13}
& 3.05\worse{$+$0.22}
& $-0.07$\worse{$-$0.28}
& 0.70\worse{$-$0.03} \\
\midrule
\multicolumn{5}{@{}l}{\textit{Correction targets}} \\
\quad inspect every element
& 1.87\worse{$+$0.12}
& 3.01\worse{$+$0.18}
& 0.13\worse{$-$0.08}
& 0.72\worse{$-$0.01} \\
\quad random (same number)
& 1.81\worse{$+$0.06}
& 2.96\worse{$+$0.13}
& 0.07\worse{$-$0.14}
& 0.71\worse{$-$0.02} \\
\bottomrule
\end{tabular}
}
\caption{Component ablations on MedleyDB. 
}
\label{tab:components}
\label{tab:selection}
\vspace{-0.1in}
\end{table}

\section{Conclusion}
\label{sec:conclusion}
We presented \method{}, an audio-language agent that mixes music by building an explicit plan of audio effects over tracks, subgroups, and the full mix, then refines it bottom-up with a learned detector and critic. Our results suggest a broader lesson for agents that act through numerical tools. Targeted, executable feedback trained on outcomes improves mixing. Because each decision logs a named audio effect and value, users can inspect the result, trace corrections, and edit the plan. 

\newpage
\section*{Limitations}
First, multiple mixes can be valid for the same input tracks, so reconstruction error against one professional mix does not fully capture mixing quality~\citep{moliner2025megami}.
In our experiments, TuneJury is useful mainly for detecting severe degradation and is not yet sufficient as a standalone objective metric for assessing mixing quality. TuneJury~\citep{kim2026tunejury} operates on mono audio (48 kHz for CLAP~\citep{wu2023large} and 24 kHz for MERT~\citep{li2024mert}), while Audiobox Aesthetics uses 16 kHz mono audio. These representations limit direct assessment of stereo imaging and, for Audiobox, high-frequency detail. Developing reliable perceptual metrics for multitrack mixing remains a separate research challenge beyond the scope of this work.
Second, the evaluation sets are small (20 MedleyDB clips from 17 songs, 14 AlbumDB clips, and 10 clips for the listening study), limiting generalizability and the precision of perceptual comparisons. Results on AlbumDB are mixed.
MST accepts only four sources, so we ran it on the separated professional mix rather than on identical input tracks, limiting direct comparability.
Third, supervision and oracles come from GRAFX-fitted parameters, which approximate, but do not recover, the engineer's settings. The large gap to GT-fitted parameters (1.75 vs.\ 0.90 MRS) shows that parameter setting remains far from solved. 
Fourth, our audio effects library covers only a small subset of the processing tools available to professional audio engineers, creating a tool gap that limits the range of achievable mixes. The plan is also fixed after planning, so the critic cannot add or remove audio effects. 
Finally, although explicit plans expose editable audio effects and parameters, we have not evaluated how users interpret or modify them in practice. Producing a mix also requires many calls to large audio-language models, limiting computational efficiency.

\section*{Ethical Considerations}
We use the MedleyDB and AlbumDB research datasets in accordance with their terms of use. Automatic mixing tools may change the work of audio engineers. We designed \method{} to expose and accept edits to any decision so that it supports, rather than replaces, human judgment. \method{} can also be used as an educational tool for audio engineers.

\bibliography{custom}

\clearpage
\appendix
\section{Experimental Setup}
\label{app:exp-setup}

\subsection{Data}
\label{app:data}
We run a music-structure analysis on each MedleyDB song and, in every section longer than 35\,s, sample one pair of non-overlapping 15-s windows at least 5\,s apart (\S\ref{sec:exp-setup}). This gives 214 candidate clips from 137 songs, of which 185 clips from 123 songs remain after filtering (44.1\,kHz, 24-bit, $-12$\,LUFS). We split by artist into 165 training clips (106 songs) and 20 evaluation clips (17 songs). The 14 AlbumDB evaluation clips are built in the same way.

\subsection{Audio effects Library}
\label{app:effects}
Table~\ref{tab:effects} lists the audio effects available to the planner
($\mathcal{T}$ in \S\ref{sec:background}), their parameters, and the levels
at which they can be used. All audio effects are differentiable processors from
GRAFX~\citep{lee2024grafx} and related libraries, so the same plan can be
rendered by the agent and fitted to GT for the oracle analyses. Every audio effect
also has a dry/wet mix control. The planner and critic write raw
(pre-activation) values. After every edit, values are clipped to a safe
range, and the ranges in Table~\ref{tab:effects} are those safe ranges in
physical units. The planner fixes the chains before parameters are set, so
refinement cannot add, remove, or reorder audio effects.

\begin{table}[!t]
\centering\small
\setlength{\tabcolsep}{3pt}
\begin{tabular}{@{}>{\raggedright\arraybackslash}p{0.2\columnwidth}>{\raggedright\arraybackslash}p{0.58\columnwidth}c@{}}
\toprule
Audio effect & Parameters (safe range) & Levels \\
\midrule
Gain / panning & left and right gain ($-7.8$ to $+2.2$\,dB each), with
panning from unequal channel gains & S, B, M \\
Parametric EQ & 10 bands (low shelf, 8 peaking, high shelf) with
frequency (0.1--10.4\,kHz), gain ($-6.1$ to $+4.3$\,dB), and $Q$
(0.45--3.3) & S, B, M \\
Compressor & threshold ($-30$ to $-12$\,dB), ratio (1.3--3.3:1), makeup
($\pm3$\,dB), attack, release, expander threshold and ratio & S, B, M \\
Multiband compressor & 3 bands (crossovers 41--279\,Hz and 2.8--11.2\,kHz),
per-band threshold ($-57$ to $-3$\,dB), ratio (1.3--17.7:1), attack,
release & M \\
Noise gate & threshold ($\approx-41$ to $-28$\,dB), ratio, knee,
smoothing & S, B, M \\
Distortion & drive (0.2--0.7\,dB) & S, B, M \\
Stereo imager & side gain ($-4.3$ to $+3.0$\,dB) & S, B, M \\
Delay & single tap with feedback, time (110--342\,ms), gain
($-44$ to $-19$\,dB), feedback, tone filter & S, B, M \\
Reverb & filtered-noise reverb, 12 bands $\times$ mid/side, per-band decay
and level & S, B, M \\
Limiter & threshold ($-44$ to $-0.2$\,dB), attack (0.5--5.4\,ms), release
(21--180\,ms) & M \\
\bottomrule
\end{tabular}
\caption{Audio effects library. S, B, and M denote track, subgroup, and full mix. Ranges are the safe ranges enforced after every edit, converted to physical units.
The multiband compressor and limiter start bypassed.}
\label{tab:effects}
\end{table}

\subsection{Baselines}
\label{app:baselines}
Single-pass baselines render their first plan. Self-Refine and Self-Critic
run up to three rounds of listening and revision, and we score the final
output. The random baseline pools valid outputs across seeds.

\subsection{Metrics}
\label{app:metrics}
\label{app:tunejury-validation}
All outputs are normalized to $-12$\,LUFS before scoring, and we report means over clips. TuneJury ranks the professional mix above randomly processed versions 70.5\% of the time, but above the dry sum only 48\% of the time and above random track gains only 41\%. It thus detects severe processing errors but not subtler differences, so we use it only as a check for severe problems.

\section{\method{} Implementation}
\label{app:method}

\subsection{Retrieval and Audio effects Chain Planning}
\label{app:planning-details}
We use frozen Qwen3-Omni~\citep{xu2025qwen3omni} for grouping and planning agents. 
Database entries store GRAFX-pruned audio effects chains~\citep{lee2024grafx,
lee2025mixinggraphs}, the mix rendered from them, and its caption. Text
queries concatenate captions of the unprocessed input tracks, generated by
Audex~\citep{kong2026audex} conditioned on instrument labels, and are
matched to database captions with Gemini
Embedding~\citep{lee2025geminiembedding}. Audio queries embed the style
reference with RLAT~\mbox{\citep{lee2026audiotransformations}} and are matched
to the rendered mixes. Both use cosine similarity, and all excerpts of the
input song are excluded. Algorithm~\ref{alg:toolplan} gives the full
planning procedure.

\begin{algorithm*}[!htbp]
\caption{Retrieval-Augmented Audio effects Chain Planning}
\label{alg:toolplan}
\begin{algorithmic}[1]
\REQUIRE unprocessed input tracks $X$ with instrument labels, reference $r$,
database $\mathcal{M}$ of mixed songs, retrieval counts
$n_{\text{text}}$ and $n_{\text{audio}}$
\ENSURE subgroup assignments and audio effects chains for all tracks, subgroups,
and the full mix, without parameters
\STATE $\{S_b\} \gets \text{\textsc{Group}}(X)$
\COMMENT{use track audio and instrument labels}
\STATE $\mathcal{M}' \gets \mathcal{M}$ without excerpts of the input song
\STATE $u_{\text{text}} \gets
\text{embed}(\text{captions of } X \text{ and its labels})$
\STATE $\mathcal{N}_{\text{text}} \gets$
top-$n_{\text{text}}$ entries of $\mathcal{M}'$
by caption similarity to $u_{\text{text}}$
\STATE $u_{\text{audio}} \gets \text{embed}(r)$
\STATE $\mathcal{N}_{\text{audio}} \gets$
top-$n_{\text{audio}}$ entries of $\mathcal{M}'$
by mix similarity to $u_{\text{audio}}$
\STATE $\mathrm{ICL} \gets$ audio effects chains of
$\mathcal{N}_{\text{text}} \cup \mathcal{N}_{\text{audio}}$
\COMMENT{parameter values omitted}
\STATE $\{f_i\} \gets
\text{\textsc{Planner}}(X, \{S_b\}, \mathrm{ICL})$
\COMMENT{track stage}
\STATE $\{f_b\} \gets
\text{\textsc{Planner}}(\{S_b\}, \{f_i\}, \mathrm{ICL})$
\COMMENT{subgroup stage using the track plans}
\STATE $f_m \gets
\text{\textsc{Planner}}(\{f_i\}, \{f_b\}, \mathrm{ICL})$
\COMMENT{full-mix stage using the track and subgroup plans}
\RETURN $\{S_b\}, \{f_i\}, \{f_b\}, f_m$
\end{algorithmic}
\end{algorithm*}

\subsection{Hierarchical Refinement}
\label{app:refinement}
Algorithm~\ref{alg:paramfill} gives the refinement procedure. \method{}
uses one pass per stage ($N=1$), and Appendix~\ref{app:iteration} studies
more passes.

\begin{algorithm*}[!htbp]
\caption{Hierarchical Refinement}
\label{alg:paramfill}
\begin{algorithmic}[1]
\REQUIRE plan $G$, tracks $X$, reference $r$,
feedback form (A1, A2, or A3), detector threshold $\tau$,
maximum passes $N$ ($N=1$ for \method{})
\ENSURE parameters $\theta$ of all elements

\STATE $\theta \gets
\text{\textsc{Planner}}(r, \text{text of } G)$

\FOR{stage $k=1,2,3$ (tracks, subgroups, full mix)}
  \FOR{pass $t=1$ \TO $N$}
    \STATE render $\mathcal{R}_k(\theta)$ and store the
    processed element audio and current mix
    \COMMENT{audio fixed throughout this pass}

    \IF{$k=3$}
      \STATE $\mathcal{Q} \gets \{m\}$
      \COMMENT{the full mix is always inspected}
      \STATE $d_m \gets \emptyset$
      \COMMENT{no detector output for the full mix}
    \ELSE
      \STATE $H \gets$ fixed audio embeddings of stored processed
      element audio and reference $r$, with audio effects chains and $\theta$
      \STATE $\{(p_e,d_e)\}_e \gets
      \text{\textsc{Detector}}(H)$
      \COMMENT{error probabilities and audio effects flags}
      \STATE $\mathcal{Q} \gets \{e \mid p_e>\tau\}$,
      highest $p_e$ first
      \IF{$\mathcal{Q}=\emptyset$}
        \STATE \textbf{break}
        \COMMENT{end passes for this stage}
      \ENDIF
    \ENDIF

    \FOR{each $e \in \mathcal{Q}$}
      \STATE $o_e \gets$ stored processed audio of $e$,
      stored mix, reference $r$, and text containing $d_e$,
      $e$'s audio effects chain in $G$, and current parameters $\theta_e$
      \STATE $a_e \sim \pi_\phi(\cdot \mid o_e)$
      \COMMENT{the critic proposes a correction}
      \STATE $\theta \gets \mathrm{Exec}(a_e,\theta)$
      \COMMENT{update only $e$, A3 directly, A1/A2 through the planner}
    \ENDFOR
  \ENDFOR

  \STATE render $\mathcal{R}_k(\theta)$
  \COMMENT{the next stage receives the corrected audio}
\ENDFOR

\RETURN $\theta$
\end{algorithmic}
\end{algorithm*}

\subsection{Detector}
\label{app:detector-arch}
The detector is a 34.9M-parameter classifier with a four-layer bidirectional Transformer encoder and two output heads. It takes fixed embeddings of each element's processed audio and of the reference $r$, together with the element's audio effects chain and current parameters. For each audio input, we average MusicFM layer 7 features~\citep{won2024musicfm} over time and concatenate them with RLAT features~\citep{lee2026audiotransformations}. One head predicts the element's error probability $p_e$, and the other identifies audio effects that may need changes. The embeddings remain fixed during detector training. The detector is trained on the 40{,}264 states in Table~\ref{tab:mixcritic-stats}. A positive label requires a replaced parameter that changes the MRS of that track by at least 1.0. Appendix~\ref{app:detector} reports its accuracy.

\subsection{Critic Architecture and Training}
\label{app:critic-training}
The critic connects a frozen RLAT audio effects encoder~\citep{lee2026audiotransformations} to Qwen3.5-4B~\citep{qwen2026qwen35} through a trained audio adapter. We use RLAT's audio effects embedding $z_T$. The three audio inputs are the selected track, subgroup, or full mix with its current audio effects, the current mix, and the reference. Each is labeled by role and encoded separately. The adapter uses 32 queries to produce 32 tokens per audio input. The text input contains the detector's output and the selected element's audio effects chain and current parameters.

SFT first trains the adapter alone (stage 1) and then trains it jointly with LoRA weights (stage 2). DPO~\citep{rafailov2023dpo} then trains on the preference pairs of \S\ref{sec:method-training},
\begin{equation}
\begin{aligned}
\mathcal{L}_{\mathrm{DPO}}(\phi)
&= -\mathbb{E}_{\mathcal{D}_{\mathrm{pref}}}
\log \sigma\!\left(
\beta [h(a^+) - h(a^-)]
\right),\\
h(a)
&= \log \pi_\phi(a \mid o)
- \log \pi_{\mathrm{SFT}}(a \mid o),
\end{aligned}
\label{eq:appendix-dpo}
\end{equation}
where $\pi_{\mathrm{SFT}}$ is the frozen SFT critic, $\sigma$ is the sigmoid function, and $\beta=0.1$. The policy starts from the SFT checkpoint, and reference log-probabilities are precomputed. We use AdamW with weight decay $0.01$. The learning rate follows 15 linear warmup steps and then cosine decay to $10^{-8}$. We use bf16 mixed precision and no gradient clipping. Table~\ref{tab:hyperparams} lists all training and inference hyperparameters.

\begin{table}[!htbp]
\centering\small
\setlength{\tabcolsep}{4pt}
\begin{tabular}{@{}>{\raggedright\arraybackslash}p{0.5\columnwidth}>{\raggedleft\arraybackslash}p{0.44\columnwidth}@{}}
\toprule
Setting & Value \\
\midrule
Critic backbone & Qwen3.5-4B + frozen RLAT \\
Audio adapter & 32 queries, 32 tokens per audio input \\
Trainable parameters (adapter + LoRA) & 89M + 99M \\
SFT learning rate (stage 1 / 2) & $10^{-4}$ / $5\cdot10^{-5}$ \\
SFT effective batch size (stage 1 / 2) & 32 / 16 \\
SFT stage 1 length & $\approx$14 epochs \\
SFT stage 2 length (selected checkpoint) & 5.8--7.4 epochs \\
LoRA rank / $\alpha$ / dropout & 128 / 256 / 0.05 \\
NTL weight & 0.3 \\
DPO $\beta$ / learning rate & 0.1 / $5\cdot10^{-6}$ \\
DPO batch size & 24 pairs \\
Margins $\delta_0$ / $\delta$ & 0.02 / 0.1 \\
DPO candidates per inspected element & 8 \\
DPO sampling (temperature, top-$p$) & 1.0, 0.95 \\
Detector threshold $\tau$ & 0.15 \\
Critic decoding (max tokens) & greedy, 3{,}072 (A3), 768 (A1/A2) \\
Planner decoding (temperature, top-$p$) & 0.6, 0.95 \\
GPUs per run & 4$\times$A100 80\,GB or 2--3$\times$H100 \\
\bottomrule
\end{tabular}
\caption{Training and inference hyperparameters.}
\label{tab:hyperparams}
\end{table}

\section{The \mixcritic{} Dataset}
\label{app:mixcritic}
Section~\ref{sec:method-training} and Figure~\ref{fig:mixcritic} describe how \mixcritic{} is built, and Table~\ref{tab:mixcritic-stats} summarizes it. All examples come from the 165 MedleyDB training clips, which share no artist with the evaluation clips. To build preference pairs, we sample candidate critiques from the SFT critic with temperature $1.0$ and top-$p$ $0.95$, and apply each valid candidate independently to the same corrupted parameters. Responses that cannot be read in the required feedback format receive a score of 0. The 2{,}200 validation examples are randomly selected from states built on validation clips, which are training clips held out from training. We select model weights on this validation set, and the evaluation clips are used only for testing.

\begin{table}[!htbp]
\centering\small
\begin{tabular}{@{}lr@{}}
\toprule
Statistic & Value \\
\midrule
Source clips / songs & 165 / 106 \\
Supervised critique examples (train, A1--A3) & 45{,}657 \\
Validation examples & 2{,}200 \\
Feedback forms & A1, A2, A3 \\
Preference pairs (DPO) & 47{,}857  \\
Detector training states & 40{,}264 \\
Corrupted tracks per state (range) & 1--36 \\
\bottomrule
\end{tabular}
\caption{\mixcritic{} statistics. Example counts are totals across the
three feedback forms.}
\label{tab:mixcritic-stats}
\end{table}

\FloatBarrier
\section{Additional Results}

\subsection{Comparison with Dedicated Mixing Models}
\label{app:dedicated}
Dedicated mixing models are trained for reference-guided mixing but differ
from our setting in their inputs or output space, so we report them
separately (Table~\ref{tab:dedicated}). Diff-MST~\citep{vanka2024diffmst}
takes the unprocessed input tracks and predicts the parameters of a fixed differentiable
console, with gain, EQ, compression, and panning on each track and EQ and
compression on the full mix. It cannot choose audio effects or group tracks.
Diff-MST is trained on MedleyDB and Cambridge-MT, so its training songs
may overlap our MedleyDB evaluation clips, and the comparison is not
strictly fair.
MST~\citep{koo2023mixingstyletransfer} is an end-to-end network that
outputs processed audio and exposes no audio effects parameters. It is trained on
four instrument classes (vocals, drums, bass, and other), and for mixture
inputs its authors first apply source separation. Following this setup, we
separate the professional mix into the four sources with
Demucs~\citep[\texttt{mdx\_extra};][]{defossez2021hybrid}, and MST
restyles each of them to match the reference. StemFX~\citep{cheng2026stemfx} predicts
variable-length audio effects chains on source-separated tracks and is applied in
the same way. MST and StemFX therefore start from the professionally mixed
version of the target, which gives them privileged access to GT.

On MedleyDB, \method{} has the lowest MSS of all four systems.
MST scores better on Audiobox PQ, FxEnc++, and AF, and StemFX on TuneJury. Both models
largely inherit their global audio effects style from the professional mix they
start from. On AlbumDB, \method{} improves
on Diff-MST in reconstruction (3.81 vs.\ 4.72 MSS), whereas MST and StemFX,
which start from the professional mix, score higher on most metrics.

\begin{table*}[!htbp]

\centering\footnotesize
\setlength{\tabcolsep}{1.5pt}
\begin{tabular*}{\textwidth}{@{\extracolsep{\fill}}lccccccccccc@{}}
\toprule
& \multicolumn{5}{c}{\textbf{MedleyDB}} &
& \multicolumn{5}{c}{\textbf{AlbumDB}} \\
\cmidrule(lr){2-6}\cmidrule(l){8-12}
& Reconst. & \multicolumn{2}{c}{Quality} & \multicolumn{2}{c}{Similarity} &
& Reconst. & \multicolumn{2}{c}{Quality} & \multicolumn{2}{c}{Similarity} \\
\cmidrule(lr){2-2}\cmidrule(lr){3-4}\cmidrule(lr){5-6}\cmidrule(lr){8-8}\cmidrule(lr){9-10}\cmidrule(l){11-12}
Method & MSS $\downarrow$ & TuneJury $\uparrow$ & Audiobox PQ $\uparrow$ & FxEnc++ $\uparrow$ & AF $\downarrow$
&& MSS $\downarrow$ & TuneJury $\uparrow$ & Audiobox PQ $\uparrow$ & FxEnc++ $\uparrow$ & AF $\downarrow$ \\
\midrule
Diff-MST & 4.31 & -0.70 & 6.49 & 0.72 & 0.19
&& 4.72 & 1.02 & 7.15 & 0.73 & 0.11 \\
\gr{MST$^\ast$} & \gr{3.08} & \gr{-0.18} & \gr{7.04} & \gr{0.79} & \gr{0.08}
&& \gr{3.60} & \gr{1.50} & \gr{7.70} & \gr{0.83} & \gr{0.04} \\
\gr{StemFX$^\ast$} & \gr{3.31} & \gr{0.28} & \gr{--} & \gr{0.65} & \gr{1.08}
&& \gr{3.55} & \gr{1.35} & \gr{--} & \gr{0.79} & \gr{0.21} \\
\midrule
\method{} & 2.83 & 0.21 & 7.03 & 0.73 & 0.20
&& 3.81 & 0.70 & 7.41 & 0.79 & 0.11 \\
\bottomrule
\end{tabular*}
\caption{Comparison with dedicated mixing models. Lower is better for MSS and AF;
higher is better for TuneJury, Audiobox PQ, and FxEnc++. Audiobox PQ is
reference-free and computed at $-12$\,LUFS.
$^\ast$Gray rows start from the professional mix separated into
four sources and thus have privileged access to GT.}
\label{tab:dedicated}
\end{table*}

\subsection{Audio effects Chain Planning}
\label{app:planning}
Table~\ref{tab:tool-planning-full} extends Table~\ref{tab:tool-planning-main}
with audio effects chain F1, which measures agreement with the pseudo-GT plan of
the professional mix. The 100 clips are drawn from MedleyDB to test
retrieval at a larger scale. This study involves no training, and all
excerpts of each query song are excluded from retrieval, so there is no
leakage. The configuration used by \method{} (Text (3) + audio (1)) raises F1 at every level, e.g., from 0.33 to 0.57 for tracks.

\begin{table*}[!htbp]
\centering
\small
\setlength{\tabcolsep}{4pt}
\begin{tabular*}{\textwidth}{@{\extracolsep{\fill}}lcccccccc@{}}
\toprule
& \multicolumn{3}{c}{Audio effects chain F1 $\uparrow$} & \multicolumn{4}{c}{GT-assisted parameter fitting} & Plan size \\
\cmidrule(lr){2-4}\cmidrule(lr){5-8}\cmidrule(l){9-9}
Retrieved examples & Track & Subgroup & Full mix & MRS $\downarrow$ & FxEnc++ $\uparrow$ & TuneJury $\uparrow$ & PQ $\uparrow$ & Tools/track \\
\midrule
No examples & 0.33 & 0.38 & 0.49 & 2.13 & 0.70 & -0.12 & 7.03 & 1.55 \\
Random (3) & 0.37 & 0.42 & 0.69 & 1.71 & 0.75 & 0.09 & 7.05 & 1.74 \\
Text (1) + audio (1) & 0.54 & 0.56 & 0.61 & 1.12 & 0.94 & 0.25 & 7.28 & 3.65 \\
Text (3) & 0.56 & 0.59 & 0.79 & 1.10 & 0.94 & 0.28 & 7.29 & 3.75 \\
Text (3) + audio (1) & 0.57 & 0.59 & 0.83 & 1.08 & 0.94 & 0.29 & 7.26 & 3.77 \\
\midrule
GRAFX (pseudo-GT plan) & 1.00 & 1.00 & 1.00 & 0.89 & 0.98 & 0.28 & 7.30 & 4.38 \\
\bottomrule
\end{tabular*}
\caption{Audio effects chain planning on 100 MedleyDB clips. Audio effects chain F1 measures agreement with the pseudo-GT plan of the professional mix. Audio scores are computed after fitting each plan's parameters to GT, so they measure what the chosen audio effects can achieve rather than agent performance. PQ is the reference-free Audiobox-Aesthetics production-quality score; TuneJury and PQ are computed at $-12$\,LUFS. GRAFX uses the pseudo-GT plan itself (F1 $=1$ by construction) and serves as an upper bound.
\method{} uses Text (3) + audio (1).}
\label{tab:tool-planning-full}
\label{tab:tool-planning}
\end{table*}

\subsection{Example Plans: \method{} and GRAFX}
\label{app:console-examples}
\begin{figure*}[!htbp]
\centering
\begin{minipage}[b]{0.49\textwidth}
  \centering
  \includegraphics[width=\linewidth]{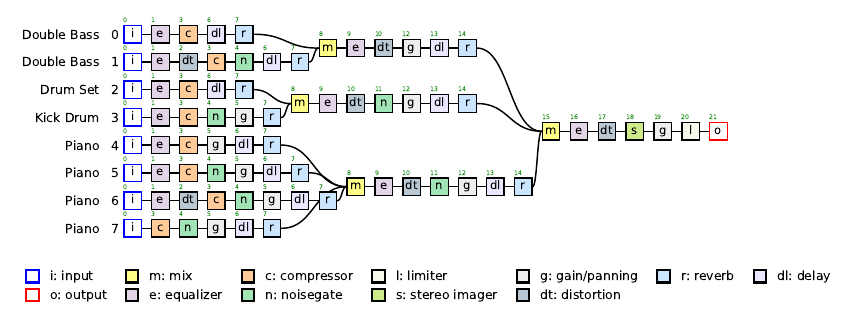}\\[1pt]
  {\small (a) GRAFX (fitted to GT, pruned)}
\end{minipage}\hfill
\begin{minipage}[b]{0.49\textwidth}
  \centering
  \includegraphics[width=\linewidth]{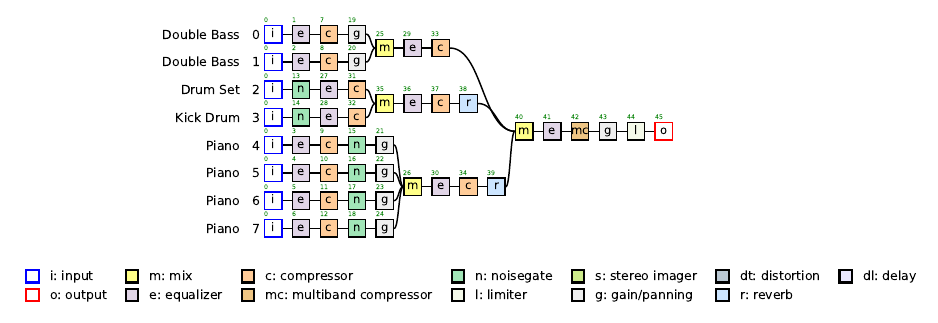}\\[1pt]
  {\small (b) \method{}}
\end{minipage}
\caption{Mixing consoles for MedleyDB clip sample\_129. Each input is labeled with its instrument and index, and input $k$ is the same track in both panels. Both consoles use the same three subgroups: double bass, drums, and piano. GRAFX is shown after pruning near-inactive processors. \method{} shows every tool in its final plan.}
\label{fig:console-medleydb-same}
\end{figure*}

\begin{figure*}[!htbp]
\centering
\begin{minipage}[b]{0.49\textwidth}
  \centering
  \includegraphics[width=\linewidth]{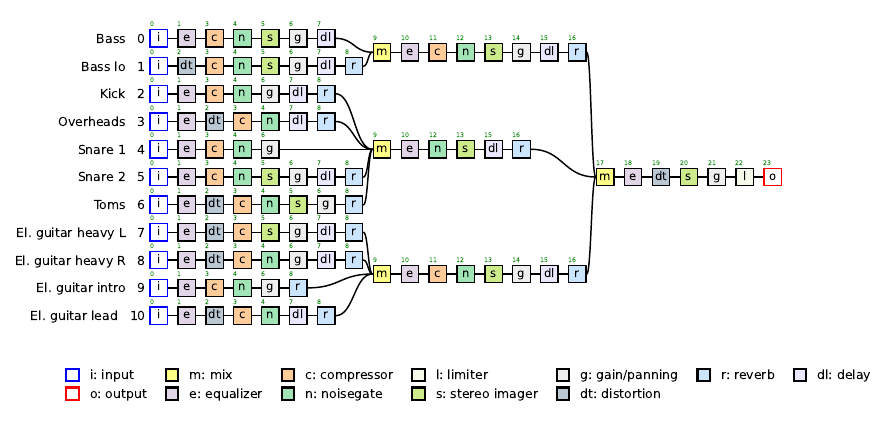}\\[1pt]
  {\small (a) GRAFX (fitted to GT, pruned)}
\end{minipage}\hfill
\begin{minipage}[b]{0.49\textwidth}
  \centering
  \includegraphics[width=\linewidth]{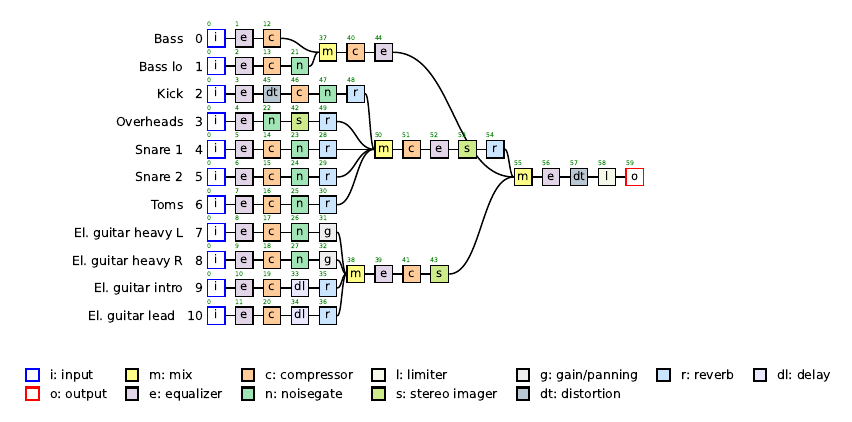}\\[1pt]
  {\small (b) \method{}}
\end{minipage}
\caption{Mixing consoles for AlbumDB clip sample\_010. Both consoles use the same three subgroups: bass, drums (kick, overheads, snares, toms), and electric guitars. Layout as in Figure~\ref{fig:console-medleydb-same}.}
\label{fig:console-albumdb-same}
\end{figure*}

\begin{figure*}[!htbp]
\centering
\begin{minipage}[b]{0.49\textwidth}
  \centering
  \includegraphics[width=\linewidth]{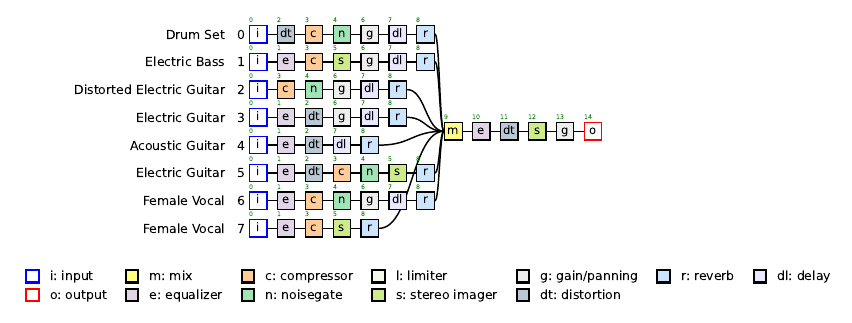}\\[1pt]
  {\small (a) GRAFX (fitted to GT, pruned)}
\end{minipage}\hfill
\begin{minipage}[b]{0.49\textwidth}
  \centering
  \includegraphics[width=\linewidth]{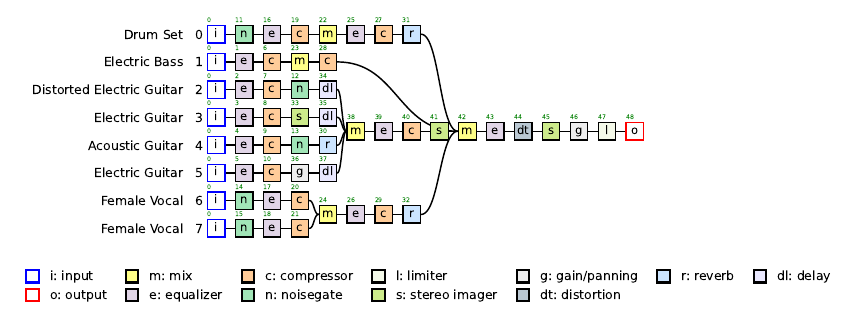}\\[1pt]
  {\small (b) \method{}}
\end{minipage}
\caption{Mixing consoles for MedleyDB clip sample\_110. The GRAFX console has no subgroups: all eight tracks feed the full mix directly. \method{} groups the four guitars and the two vocals into subgroups and gives drums and bass their own subgroups. Layout as in Figure~\ref{fig:console-medleydb-same}.}
\label{fig:console-medleydb-diff}
\end{figure*}

\begin{figure*}[!htbp]
\centering
\begin{minipage}[b]{0.49\textwidth}
  \centering
  \includegraphics[width=\linewidth]{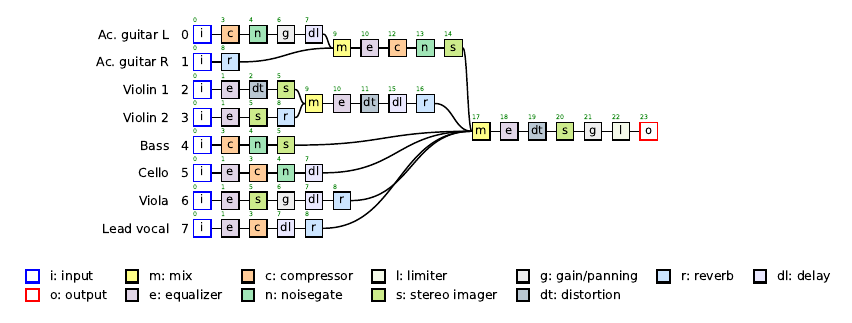}\\[1pt]
  {\small (a) GRAFX (fitted to GT, pruned)}
\end{minipage}\hfill
\begin{minipage}[b]{0.49\textwidth}
  \centering
  \includegraphics[width=\linewidth]{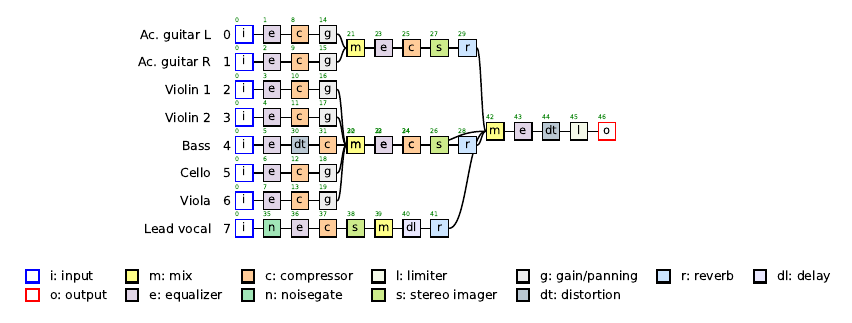}\\[1pt]
  {\small (b) \method{}}
\end{minipage}
\caption{Mixing consoles for AlbumDB clip sample\_013. The GRAFX console groups the two acoustic guitars and the two violins, and sends bass, cello, viola, and lead vocal to the full mix directly. \method{} places the two violins, cello, viola, and bass in one subgroup, the two acoustic guitars on another, and the lead vocal in its own subgroup. Layout as in Figure~\ref{fig:console-medleydb-same}.}
\label{fig:console-albumdb-diff}
\end{figure*}

\method{} builds its plan the way an engineer would. Its grouping agent
listens to the tracks and reads their labels before assigning subgroups, and its
planner chooses the audio effects and their order for each track, subgroup, and the
full mix. GRAFX, which builds the pseudo-GT plans used for retrieval and
audio effects chain F1 (\S\ref{sec:method-toolplan}), instead starts from a fixed
console that applies every audio effect in the same order everywhere. It fits all
parameters to the professional mix and prunes processors that barely affect
the result~\citep{lee2024grafx,lee2025mixinggraphs}.
Figures~\ref{fig:console-medleydb-same}--\ref{fig:console-albumdb-diff}
compare the final plans of \method{} with GRAFX plans for two MedleyDB and
two AlbumDB clips.

\paragraph{Grouping.}
\method{} forms subgroups that follow the instrumentation. In
Figure~\ref{fig:console-medleydb-diff}, it creates drum, bass, guitar, and
vocal subgroups, whereas GRAFX uses no subgroups and sends all eight tracks to the
full mix. In Figure~\ref{fig:console-albumdb-diff}, it places the violins,
cello, and viola in one subgroup together with the bass, whereas GRAFX groups
only the two violins and splits the rest of the string section. In
Figures~\ref{fig:console-medleydb-same} and~\ref{fig:console-albumdb-same},
both plans use the same subgroups.

\paragraph{Audio effects order.}
\method{} chooses the order for each element. It places the noise gate
before the compressor on the drums (Figure~\ref{fig:console-medleydb-same})
and on the vocals (Figures~\ref{fig:console-medleydb-diff}
and~\ref{fig:console-albumdb-diff}), as engineers usually do, and it
compresses before equalizing on the AlbumDB bass and drum subgroups
(Figure~\ref{fig:console-albumdb-same}). Every GRAFX chain keeps the console
order of EQ, distortion, compressor, noise gate, stereo imager,
gain/panning, delay, and reverb, so its gate always follows the compressor.

\paragraph{Audio effects choice.}
\method{} uses two to five audio effects per track, adds no delay or reverb to bass
tracks, and often applies reverb once on a subgroup rather than on each track, as
on the drum and piano subgroups in Figure~\ref{fig:console-medleydb-same}. Its
full-mix chain varies across clips and includes a multiband compressor in
Figure~\ref{fig:console-medleydb-same}. GRAFX keeps one to seven audio effects per
track and adds delay or reverb to most tracks, including the bass tracks in
Figures~\ref{fig:console-medleydb-same} and~\ref{fig:console-albumdb-same}.
Its full-mix chain always contains EQ, distortion, a stereo imager, and
gain/panning, and three of the four add a limiter.

Because audio effects chain F1 (Table~\ref{tab:tool-planning-full}) measures
agreement with GRAFX plans, it counts such choices as errors when they
differ from GRAFX. We therefore also report audio scores after fitting each
plan's parameters.

\FloatBarrier

\subsection{Detector Accuracy}
\label{app:detector}
On a validation set of 3{,}130 corrupted states from 13 training clips
held out from training, the detector
reaches 0.81 F1 for tracks and 0.77 for audio effects at threshold 0.5
(Table~\ref{tab:detector-validation}). This threshold is used only for
validation, and \method{} uses $\tau=0.15$ at inference.

\begin{table}[!htbp]
\centering\small
\setlength{\tabcolsep}{3pt}
\begin{tabular*}{\columnwidth}{@{\extracolsep{\fill}}lccccc@{}}
\toprule
Target & $N$ & Prec. & Recall & F1 & AUROC \\
\midrule
Track & 9,779 & 0.797 & 0.822 & 0.809 & 0.893 \\
Audio effect & 33,329 & 0.700 & 0.849 & 0.767 & 0.874 \\
\bottomrule
\end{tabular*}
\caption{Detector validation at threshold 0.50. $N$ counts decisions.}
\label{tab:detector-validation}
\end{table}

\subsection{Error at Each Stage}
\label{app:stage-error}
Table~\ref{tab:stage_error} measures MRS after each stage against the
GT-fitted audio of that stage. With numerical feedback, DPO has lower error
than SFT at every stage.

\begin{table}[!htbp]
\centering\small
\setlength{\tabcolsep}{4pt}
\begin{tabular*}{\columnwidth}{@{\extracolsep{\fill}}lcccc@{}}
\toprule
Critic & Track & Subgroup & Full mix & Final \\
\midrule
Untrained A1 & 1.50 & 1.82 & 2.12 & 2.26 \\
SFT A1 & 1.46 & 1.73 & 1.99 & 2.08 \\
DPO A1 & 1.47 & 1.81 & 1.96 & 2.09 \\
\midrule
Untrained A2 & 1.71 & 2.13 & 2.34 & 2.44 \\
SFT A2 & 1.53 & 1.97 & 2.19 & 2.24 \\
DPO A2 & 1.60 & 1.88 & 1.89 & 1.98 \\
\midrule
Untrained A3 & 1.46 & 1.81 & 2.09 & 2.18 \\
SFT A3 & 1.42 & 1.88 & 1.84 & 1.91 \\
DPO A3 & 1.33 & 1.80 & 1.68 & 1.75 \\
\bottomrule
\end{tabular*}
\caption{MRS after each processing stage (lower is better). Track, subgroup, and full mix targets
are the corresponding GT-fitted outputs. The final target is the real mix.
Untrained rows use prompted Qwen3-Omni as the critic.}
\label{tab:stage_error}
\end{table}

\subsection{Number of Refinement Passes}
\label{app:iteration}
\begin{figure*}[!htbp]
\centering
\includegraphics[width=0.328\textwidth]{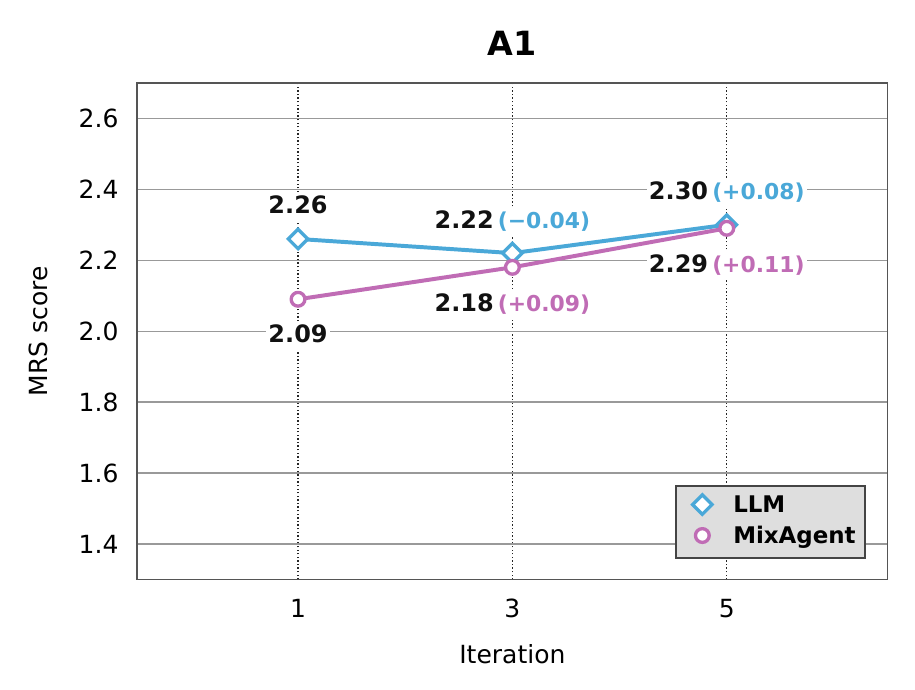}\hfill
\includegraphics[width=0.328\textwidth]{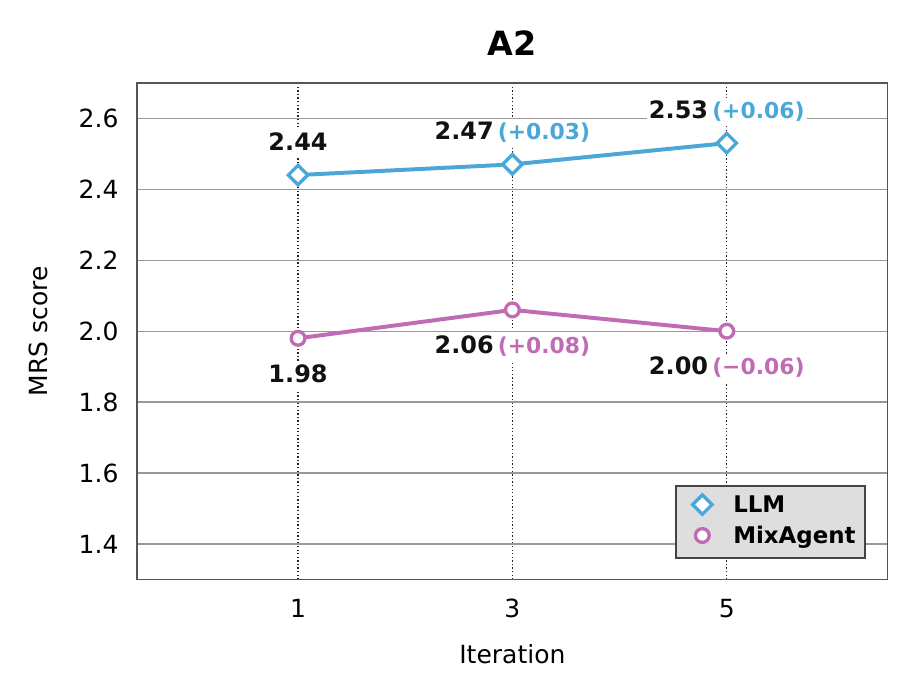}\hfill
\includegraphics[width=0.328\textwidth]{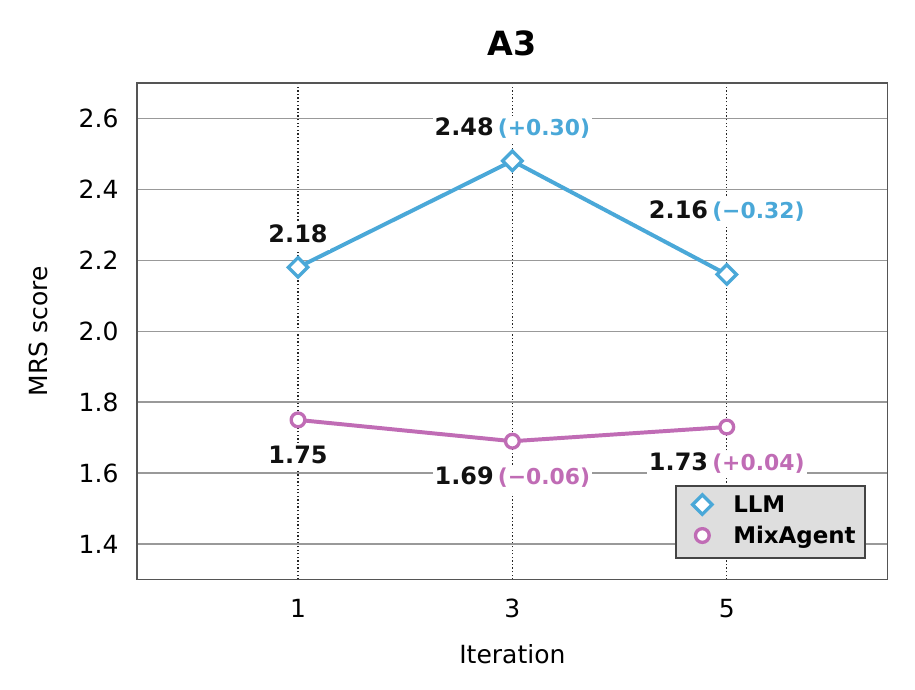}
\caption{MRS (lower is better) with a maximum of one, three, or five
refinement passes per stage for A1, A2, and A3, from left to right. Parentheses
show the score change from the preceding plotted pass budget.}
\label{fig:iteration}
\end{figure*}

Additional passes do not consistently help (Figure~\ref{fig:iteration}).
With A3, \method{} improves from 1.75 to 1.69 MRS with three passes and
returns to 1.73 with five, while A1 and A2 are best with one pass.

\FloatBarrier
\section{Ablation Protocols}
\label{app:ablation-protocol}

\subsection{Critic Training and Feedback Form}
\label{app:critic-comparison}
\label{app:revision-scores}
All rows use the same learned detector. The untrained rows prompt
Qwen3-Omni as the critic. The trained critic is a different model (Table~\ref{tab:hyperparams}), and DPO
starts from its SFT checkpoint. Within each
feedback form, SFT and SFT+DPO share the detector, planner, plan, initial
state, prompts, and decoding. Later states can diverge as corrections are
applied. For each applied correction,
$\Delta\mathrm{MRS}=\mathrm{MRS}_{\mathrm{before}}-\mathrm{MRS}_{\mathrm{after}}$
on the full mix, and a correction counts as an improvement if
$\Delta\mathrm{MRS}>0$. Oracle A1 and A2 replace the critic's output with
the GT critique in the same feedback form, which the planner converts into
parameter values as in the other A1 and A2 rows. Oracle A3 fits the
parameters of the same plan directly to GT with GRAFX.

\subsection{Target Selection}
\label{app:selection-protocol}
\label{sec:exp-selection}
All conditions share the plan, starting state, DPO A3 critic, and the
detector's audio effects-level diagnoses, and they visit selected elements in
order of detector score. Only the set of selected tracks and subgroups changes.
The full mix is always inspected. Inspect-all averages 13.1 turns and
56.6 minutes per clip, and random selection 12.1 turns.

\subsection{Refinement Order}
\label{app:hierarchy-protocol}
Top-down visits the full mix, then subgroups, then tracks, rendering between
levels (11.9 turns and 54.2 minutes per clip). The variant without
hierarchy computes all corrections from the initial audio and applies them
at once (11.8 turns).
\FloatBarrier

\section{Prompts}
\label{app:prompts}
\makeatletter\setlength{\@dblfptop}{0pt}\setlength{\@dblfpsep}{12pt}\setlength{\@dblfpbot}{0pt plus 1fil}\makeatother
Figures~\ref{fig:prompt-grouping}--\ref{fig:prompt-critic-formats} show the
prompts of the language-model calls in \method{} and the prompt used to build \mixcritic{}. Text in braces, such as
\texttt{\{stem\_count\}}, is filled in at run time, and ``[...]'' marks lines
cut for space.

\lstdefinestyle{promptstyle}{basicstyle=\ttfamily\scriptsize, columns=fullflexible,
  keepspaces=true, breaklines=true, showstringspaces=false, aboveskip=0pt, belowskip=0pt}
\newtcblisting{prompt}[1]{enhanced, listing only,
  listing options={style=promptstyle},
  colback=gray!5, colframe=gray!45, boxrule=0.4pt, arc=1pt,
  left=4pt, right=4pt, top=3pt, bottom=3pt,
  colbacktitle=gray!18, coltitle=black, fonttitle=\footnotesize\bfseries, title={#1}}

\begin{figure*}[!htbp]
\begin{prompt}{Grouping agent}
You are an expert mixing engineer setting up a session. You have direct audio access to {stem_count} full dry stem audio files (attached in fixed order) plus their filenames. You CAN hear every file -- listen carefully before you respond.

Dry stem filename order:
{stem_file_list}

Your task is ONLY to group these {stem_count} dry stems into a small number of mixing buses (submix groups), exactly as an engineer routes related tracks to shared buses before balancing them. Do NOT choose any tools or parameters yet and do NOT try to mix -- only decide the grouping.

Grouping guidance:
- Infer each stem's role from its filename and audio content: drums (kick, snare, toms, hats, overheads, room), bass / low-end foundation, lead vocal, backing/harmony vocals, [...]
- Put stems that should be balanced and processed together into the same group. For example, all drum mics belong in one "drums" bus; a lead vocal with its doubles and harmonies belongs in one "vocals" bus; the DI and amp of a single guitar belong together.
- Choose the number of groups yourself based on what the song actually contains. Typical mixes use about 3 to 6 groups. Use at most {max_groups} groups.
- Every stem must be placed in exactly one group. [...]
- Prefer a few meaningful buses over many tiny ones. A solo instrument with no relatives can be its own group.

Output format. Return a single JSON object with exactly one key "groups": [...]
{"groups": [{"name": "drums", "stem_indices": [1, 2, 3]},
            {"name": "vocals", "stem_indices": [4, 5]},
            {"name": "bass", "stem_indices": [6]}]}
\end{prompt}
\caption{Prompt of the grouping agent, which hears all unprocessed input tracks together with their file names. The reproduced prompts retain their original terminology: stems are input tracks, buses process subgroups, and the master bus processes the full mix.}
\label{fig:prompt-grouping}
\end{figure*}

\begin{figure*}[!htbp]
\begin{prompt}{Tool-planning agent: stem and bus steps}
You are an expert mixing engineer planning ONE submix bus ("{group_name}") of a song. You have direct audio access to the {stem_count} dry stem audio files of this bus, attached in fixed order. You CAN hear every file -- listen carefully before you respond.

Bus stem filename order:
{stem_file_list}

Your task: choose WHICH mixing tools to apply -- you are NOT choosing any parameter values in this pass. A separate optimizer will set all parameters later; you only decide the processing structure, exactly like an engineer patching processors into a console before touching any knob.

Decide two things:
1. For each stem in this bus: an ordered per-stem tool chain (possibly empty, if the stem needs no individual processing before the bus).
2. One ordered tool chain for the bus itself (the shared "{group_name}" group chain applied to the summed bus; possibly empty).

Allowed tools (both per-stem and bus chains): eq, distortion, compressor, noisegate, stereo_imager, gain_panning, delay, reverb.

Guidance:
- Choose tools by role and by what you hear: gates for bleed, compressors for uneven dynamics, eq for tone/separation, gain_panning for level/placement, reverb/delay for space, stereo_imager for width, distortion for grit.
- Don't default to minimal. A single tool per stem is usually too little [...]
- The in-context examples show the kind of fuller, multi-tool treatment expert mixes use. Treat them as an illustration of how thorough to be when a stem calls for it, not a tool count to copy.
- Put shared, glue-style processing (bus compression, shared ambience, overall bus tone) in the group chain -- but still give each individual stem its own full treatment [...]
[...]

{icl_examples}

Output format. Return a single JSON object with exactly these two keys: [...]
{"stems": [{"stem_index": 1, "chain": ["noisegate", "eq", "compressor", "gain_panning"]},
           {"stem_index": 2, "chain": ["eq", "compressor"]}],
 "group_chain": ["compressor", "eq", "gain_panning"]}
\end{prompt}
\caption{Prompt of the tool-planning agent for the track and subgroup steps (one call per subgroup). The retrieved examples replace \texttt{\{icl\_examples\}} (Figure~\ref{fig:prompt-icl}).}
\label{fig:prompt-tools-bus}
\end{figure*}

\begin{figure*}[!htbp]
\begin{prompt}{Tool-planning agent: master step}
You are an expert mixing engineer finalizing the master bus of a song. You choose WHICH tools to apply on the master bus -- no parameter values; a separate optimizer will set all parameters later.

Session structure (groups, their member stems, and the tool chains already chosen):
{session_summary}

Your task: one ordered tool chain for the master bus (applied to the sum of all group buses and passthrough stems).

Allowed master-bus tools: eq, distortion, compressor, noisegate, stereo_imager, gain_panning, delay, reverb, multiband_compressor, limiter.

Guidance:
- Typical masters use a handful of tools: tonal eq, glue compression (single-band or multiband), optional width/saturation, a level trim, and a final limiter last when a peak ceiling is needed.
- Choose based on what the session structure implies (how much dynamics/tone work already happened at stem/group level) and general mastering practice. Empty is allowed but rare.

{icl_examples}

Output format. Return a single JSON object with exactly one key:
{"master_chain": ["eq", "compressor", "gain_panning", "limiter"]}
\end{prompt}
\caption{Prompt of the tool-planning agent for the full mix step. This call is text only, and \texttt{\{session\_summary\}} lists the subgroups, their tracks, and the chains already chosen.}
\label{fig:prompt-tools-master}
\end{figure*}

\begin{figure*}[!htbp]
\begin{prompt}{Retrieved examples (replaces \{icl\_examples\})}
In-context examples from previously mixed songs, structured by an expert hierarchical mixing console (a gradient-descent optimizer). Guidance only -- adapt to THIS song; do not copy blindly. Parameters are intentionally omitted: study WHICH tools were kept per stem, per bus, and on the master, and how selective the choices are. [...]

Example [text] (text 0.901) -- 16 stems, 4 buses, 0 routed straight to master:
  Bus 'piano'  bus chain: eq -> compressor -> noisegate
     1  piano        eq -> compressor -> gain_panning -> delay -> reverb
        sounds like: An isolated dry piano stem with no other instruments or ambience. [...]
     2  piano        eq -> compressor -> noisegate -> delay -> reverb
        sounds like: [...]
  Bus 'double bass'  bus chain: eq -> compressor -> stereo_imager
     5  double bass  eq -> compressor -> gain_panning
        sounds like: [...]
  [...]
  Master chain: eq -> distortion -> gain_panning

[...]
\end{prompt}
\caption{Format of the retrieved examples, shown for one example
retrieved for MedleyDB clip sample\_129. \method{} inserts three examples
retrieved by text and one retrieved by audio.}
\label{fig:prompt-icl}
\end{figure*}

\begin{figure*}[!htbp]
\begin{prompt}{Planner: parameter initialization (stem step)}
You are a professional mixing engineer. You are mixing ONE instrument group of song {song_id}: the "{group_name}" group. The tools on each stem are already chosen and FIXED -- your job is to set their PARAMETERS so this group sounds right and sits the way the same group sits inside the Reference mix.

## AUDIO ORDER
{audio_order}

Listening notes:
- Each dry stem clip is loudness-normalized on its own, so the stem clips do NOT tell you the current relative levels. Set relative levels yourself with the gain_panning sites (where present) and compressor make_up.
- Match the Reference closely: how loud, bright, wide, dry/wet and dynamic is this group inside the Reference? That is your target.
- The dry stems are MONO and all sit dead-center. The mix can ONLY become stereo through UNEQUAL [left, right] values on gain_panning sites [...]

## THE GROUP'S STEMS
{group_table}

## PARAMETER SITES (set values ONLY for these numbered sites)
{site_table}

## TOOL PARAMETER CONTRACT
{tool_contract}

## YOUR TASK
Listen to every stem, then choose parameters for the sites above so that, summed together, the group is balanced internally (lead elements clear, doubles/mics supporting, no masking), panned into a sensible stereo picture where gain_panning exists, tonally clean (cut mud, tame harshness -- cuts before boosts), and dynamically controlled but alive. [...] Restraint wins: near-neutral parameters on most sites plus a few decisive moves beat processing everything.

## OUTPUT
Return ONLY one json fenced block:
{"sites": {"<site_id>": {"<param>": <value>, ...}, ...}, "notes": "<one sentence of intent>"}
[...]
\end{prompt}
\caption{Prompt for parameter initialization by the planner (track step). The subgroup and full mix steps use analogous prompts. \texttt{\{tool\_contract\}} gives each tool's parameter names, units, and valid ranges.}
\label{fig:prompt-init}
\end{figure*}

\begin{figure*}[!htbp]
\begin{prompt}{\method{} critic (A3)}
You are a mixing critic for one element of a multitrack music mix.

You will hear three audio clips, in this order:
1. TARGET -- {target_clip_desc}
2. MIX -- the current mix.
3. REFERENCE -- the reference mix for this song: how it should sound.

The detector flagged these elements as sounding wrong, worst first: {flagged_list}.

Criticize ONLY the target element: {target_name} ({target_instrument}, {target_level} level).
The other flagged elements are context -- do not propose fixes for them.

Diagnosis (given): the corrupted tools on this element are: {diagnosis_tools}. Exactly these tools are set wrong; every other tool in the chain is already correct.

{chain_header}
{chain_block}

Listen and compare the clips, think it through, then answer with ONE JSON object only:
{"tools": [{"tool": "<a tool name from the chain above>", "comment": "apply the parameter values below", "parameters": {<the tool's new raw parameter values>}, "drywet": <new raw drywet value>}, ...],
 "keep": "<one clause: what already works and must be preserved>"}
Give exactly one entry for each corrupted tool listed above, carrying the exact new values.
\end{prompt}
\caption{Prompt of the \method{} critic at inference, which uses the A3 answer format. Each audio clip is preceded by a text label. \texttt{\{diagnosis\_tools\}} are the audio effects flagged by the detector, and \texttt{\{chain\_block\}} lists the element's audio effects with their current values. The critic reasons before answering, and only the final JSON is used.}
\label{fig:prompt-critic}
\end{figure*}

\begin{figure*}[!htbp]
\begin{prompt}{\mixcritic{} construction: critique prompt (A1, A2, A3)}
You are a mixing critic for one element of a multitrack music mix. You do not hear the audio. Judge the element from its effect settings.

Target element: {target_name} ({target_instrument}, {target_level} level).
Other elements set wrong in this mix: {flagged_list}. They are context only -- do not propose fixes for them.

Diagnosis (given): the corrupted tools on this element are: {diagnosis_tools}. Exactly these tools are set wrong; every other tool in the chain is already correct.

The target element's effect chain, in raw parameter space. Each tool lists its CURRENT values. Each corrupted tool also lists its CORRECT values, the settings that fix the problem:
{chain_block}
  e.g.  eq         CURRENT {...}                      (already correct)
        noisegate  CURRENT {...}   CORRECT {...}      (corrupted)

Compare the CURRENT and CORRECT values. Explain what would sound wrong in the target element with the CURRENT values and how changing the corrupted tools to their CORRECT values fixes it. Then answer with ONE JSON object only:
{answer_format}
\end{prompt}
\caption{Prompt used to build \mixcritic{}, shared by A1, A2, and A3. The teacher LLM receives text only. The answer format of each feedback form replaces \texttt{\{answer\_format\}} (Figure~\ref{fig:prompt-critic-formats}). \texttt{\{diagnosis\_tools\}} are the corrupted audio effects, and \texttt{\{chain\_block\}} lists the element's audio effects with their current values and, for each corrupted audio effect, its pseudo-GT values $\theta^\ast$.}
\label{fig:prompt-critic-data}
\end{figure*}

\begin{figure*}[!htbp]
\begin{prompt}{Critic answer formats (replaces \{answer\_format\})}
A1 (free text):
{"feedback": "<2-4 sentences of plain-language criticism of the target element -- what sounds wrong and how it should sound instead. No effect names, no parameter names, no numbers.>"}

A2 (per-tool comments):
{"tools": [{"tool": "<a tool name from the chain above>", "comment": "<1-3 sentences: what is wrong involving this tool and how to fix it. No numbers.>"}, ...],
 "keep": "<one clause: what already works and must be preserved>"}
Give exactly one entry for each corrupted tool listed above.

A3 (parameter values):
{"tools": [{"tool": "<a tool name from the chain above>", "comment": "apply the parameter values below", "parameters": {<the tool's new raw parameter values>}, "drywet": <new raw drywet value>}, ...],
 "keep": "<one clause: what already works and must be preserved>"}
Give exactly one entry for each corrupted tool listed above, carrying the exact new values.
\end{prompt}
\caption{Answer formats of A1, A2, and A3, which replace \texttt{\{answer\_format\}} in Figure~\ref{fig:prompt-critic-data}.}
\label{fig:prompt-critic-formats}
\end{figure*}

\FloatBarrier

\end{document}